\documentclass[pra,reprint,superscriptaddress]{revtex4-2}

\usepackage{graphicx}
\usepackage{enumitem}
\usepackage{amsmath}
\newcommand{\ket}[1]{|{#1}\rangle}
\newcommand{\bra}[1]{\langle{#1}|}

\usepackage{amsfonts}

\usepackage[usenames,dvipsnames]{xcolor}
\usepackage[colorlinks=true,citecolor=Cerulean,linkcolor=RubineRed,urlcolor=Cerulean]{hyperref}

\usepackage{placeins}

\begin{document}

\title{Tensor network methods for the Gross-Pitaevskii equation on fine grids}

\author{Ryan J. J. Connor}
\email{ryan.connor.2019@uni.strath.ac.uk}
\affiliation{Department of Physics and SUPA, University of Strathclyde, Glasgow G4 0NG, United Kingdom}
\author{Callum W. Duncan}
\affiliation{Department of Physics and SUPA, University of Strathclyde, Glasgow G4 0NG, United Kingdom}
\affiliation{Aegiq Ltd., Cooper Buildings, Arundel Street, Sheffield, S1 2NS, United Kingdom}
\author{Andrew J. Daley}
\affiliation{Department of Physics, University of Oxford, Clarendon Laboratory, OX1 3PU Oxford, UK}
\affiliation{Department of Physics and SUPA, University of Strathclyde, Glasgow G4 0NG, United Kingdom}

\date{\today}

\begin{abstract}
The Gross-Pitaevskii equation and its generalisations to dissipative and dipolar gases have been very useful in describing dynamics of cold atomic gases, as well as polaritons and other nonlinear systems. For some of these applications the numerically accessible grid spacing can become a limiting factor, especially in describing turbulent dynamics and short-range effects of dipole-dipole interactions.  We explore the application of tensor networks to these systems, where (in analogy to related work in fluid and plasma dynamics), they allow for physically motivated data compression that makes simulations possible on large spatial grids which would be unfeasible with direct numerical simulations. Analysing different non-equilibrium cases involving vortex formation, we find that these methods are particularly efficient, especially in combination with a matrix product operator representation of the quantum Fourier transform, which enables a spectral approach to calculation of both equilibrium states and time-dependent dynamics. The efficiency of these methods has interesting physical implications for the structure in the states that are generated by these dynamics, and provides a path to describe cold gas experiments that are challenging for existing methods.

\end{abstract}

\maketitle

\section*{Introduction}

Tensor networks are an an indispensable tool used in the study of quantum many-body physics \cite{biamonte2017tensornetworksnutshell,TN_quantum_systems,Or_s_2014}, allowing for the study of low lying energy states of many-body Hamiltonians, beginning with the Density Matrix Renormalisation Group \cite{Schollw_ck_2011,chan2016matrixproductoperatorsmatrix}, and allowing for the study of out of equilibrium physics via time evolution \cite{Paeckel_2019,Vanderstraeten_2019,TDVP,Schollw_ck_2011}. Tensor networks can allow for the study of large systems consisting of many particles by systematic data compression in the sense of reducing the number of parameters that must be stored, provided that the amount of entanglement present in the states they represent is appropriately restricted \cite{Evenbly_2011,cichocki2014tensornetworksbigdata,schuch2008}. Furthermore, previous work has explored the application of tensor networks to the simulations of continuous quantum systems and beyond for applications to solutions of linear equations and even Poisson like differential equations \cite{PhysRevB.75.104305, PhysRevB.105.165116}.

Recent work has begun to explore the possibility of applying tensor networks to non-linear partial differential equations (PDEs), such as modelling the motion of fluids and plasmas \cite{Q_inspired_fluids,Wall_flows,Plasma_MPS,multigrid,Plasma_comb,gourianov2024tensornetworksenablecalculation,holscher2024quantuminspiredfluidsimulation2d}. Numerically simulating non-linear PDEs remains a challenging task for problems of importance in these areas, especially when there is a substantial range of scales relevant to the dynamics. Solving PDEs with Direct Numerical Simulation (DNS) involves discretising the solution space on which the quantity of interest is specified, at the expense of a computational cost that scales with the number of grid points. In contrast, TN approaches allow us to discard correlations which do not contribute to the overall physics, exploiting underlying correlation structures of the problem at hand, and the computational complexity can scale logarithmically with, e.g., the number of grid points, provided there is appropriate structure in the solution (specifically, limits in the types of correlations present).
 
Non-linearity appears in many physical systems, and this includes a variety of non-linear Schr\"odinger equations, e.g., in nonlinear optics. The theoretical description of the dynamics of cold gases, both with contact interactions and dipolar interactions are described by the Gross-Pitaevskii equation (GPE) \cite{PethickChristopher2008Bcid, pitaevskii2016bose,FINESS_book,doi:10.1080/00018730802564254}. The GPE and its extended versions have been useful for some time in describing experimentally observed dynamics and phenomena such as vortex formation in cold atomic gases. There is still significant ongoing interest in the GPE, especially in regards to quantum turbulence \cite{Ashton_turbulence,KOBAYASHI2021107579}, formation of quantised vortices \cite{Giant_Vortex_Gauthier_2019,Baggaley_2018,Hern_ndez_Rajkov_2024} and exotic properties in dipolar gases \cite{poli2024excitationstwodimensionalsupersolid,PhysRevA.108.053321,Bland_2024,PhysRevA.109.023313}, with many experimental groups currently involved in exploring such phenomena \cite{Gauthier_2019,Sachkou_2019,Yshin,Navon_2019,Norcia_2021,Tanzi_2021,Cabrera_2018,dipolar_molecules}.

There remain certain challenges in performing numerical simulations with the GPE however, such as being able to capture vastly different length scales within a simulation. To perform a physically accurate simulation one must be able to resolve the smallest size details present in the dynamics, which for turbulent dynamics, or systems with anisotopic short-range interactions, may be orders of magnitudes smaller than the size of the simulation domain. The memory requirements for performing a simulation over such large separations in length scales becomes a bottleneck, restricting the possible spatial resolution. As such, a natural question remains as to whether tensor networks can be used to provide data compression to overcome these current limitations.

In this work we extend the application of tensor networks, in the form of Matrix Product States, to perform simulations of quantum fluids and turbulence through the GPE. We exploit novel techniques such as the implementation of the quantum Fourier transform \cite{QFT} in matrix product operators to explore and compare various methods for studying dynamics of a trapped BEC in experimentally relevant settings.  The remainder of this paper is structured as follows: In Section 1, we outline the details of the GPE, before introducing tensor networks and discussing how one can use them to encode the GPE. We then outline how one can implement the commonly used split-step Fourier method with tensor networks in Section 2, which are then illustrated by examples simulations in Section 3. We highlight the use of tensor networks for modelling simple Soliton propagation, and for a more challenging test case of vortex formation in 2D and 3D.  Finally, in Section 4 we demonstrate how a dipolar gas can equally be simulated with tensor networks.

\section{Simulating the Gross-Pitaevskii equation with tensor networks}

\subsection{The Gross-Pitaevskii equation}
 
The Gross-Pitaevskii equation (GPE) describes the dynamics of a dilute Bose gas via the classical field $\Psi(\mathbf{r};t)$  \cite{PethickChristopher2008Bcid, pitaevskii2016bose} 
\begin{equation}
    i\hbar \frac{\partial \Psi(\mathbf{r};t)}{\partial t} = \left[ -\frac{\hbar^2}{2m} \nabla ^2 + V(\mathbf{r}) -\mu + g\left| \Psi(\mathbf{r};t)\right|^2  \right]\Psi(\mathbf{r};t) , \label{eq:GPE}
\end{equation}
with interaction parameter $g$, atomic mass $m$, chemical potential $\mu$, and a trapping potential $V(\mathbf{r})$. The non-linear term of the GPE is due to the assumption of a soft s-wave scattering of the atoms and appears in the typical self-consistent form of mean field equations \cite{pitaevskii2016bose}. The GPE is often a good approximation for the dynamics of a BEC in the weakly interacting regime, and also takes the same form as the non-linear Schrodinger equation \cite{sulem2007nonlinear}.

The GPE as outlined in Eq.~\eqref{eq:GPE} does not include any inherent mechanism for damping excitations, which is known to be present in a dilute BEC \cite{mewes1996collective,jin1997temperature}. We will consider the introduction of a phenomenological damping term to give the modified GPE \cite{choi1998phenomenological,pitaevskii1959phenomenological}
\begin{equation}\label{eq:GPEmodified}
    i\hbar \frac{\partial \Psi}{\partial t} = (1-i \gamma)\left[ -\frac{\hbar^2}{2m} \nabla ^2 + V(\mathbf{r}) -\mu + g|\Psi|^2  \right]\Psi ,
\end{equation}
with $\gamma$ the dissipation parameter. Note that we have dropped the explicit space and time dependence of the state for ease of reading in Eq.~\eqref{eq:GPEmodified}. Throughout this paper we use non-dimensional units, where we set $\hbar =1$ and $m=1$.

\subsection{Representing the state as a tensor network} \label{Represeting as a TN}
We will begin by considering the representation of $\Psi(\mathbf{r};t)$ as a tensor network. In this work we will focus on the case of encoding the state into a matrix product state (MPS) (which is a specific 1D tensor network, and sometimes referred to as a tensor train). We note that other geometries could be considered for encoding the problem onto tensor networks and might better representat correlations in some scenarios (though often at the cost of how the algorithm scales with the size of the tensor network). This has been investigated, e.g., for PDEs from Plasma dynamics \cite{Plasma_comb}.

To encode $\Psi$ onto an MPS we need to discretise the $d$-dimensional state $\Psi(\mathbf{r};t)$ into a grid of $2^{dN}$ gridpoints $\mathbf{r}_q$. This is equivalent to that for direct numerical approaches that do not utilise the tensor network formalism. We will take the gridpoint index $q^i$ to be defined via $N$ bits
\begin{equation}
    q^i = \left( \sigma_1^i, \sigma_2^i, \dots, \sigma_N^i \right),
\end{equation}
with $i$ the dimension ($ i \in \{ 1,d \}$) and $\sigma_j^i \in \{ 0,1 \}$. We can then write the discretised $\Psi$ as a tensor
\begin{equation}
    \Psi(\mathbf{r}_q;t) = A_{\Pi_{i=1}^d  \sigma_1^i, \dots, \sigma_N^i},
\end{equation}
with $A_{\Pi_{i=1}^d \sigma_1^i, \dots, \sigma_N^i}$ being a $dN$-order tensor. We can then apply the approach of singular value decomposition (SVD) on the $dN$-order tensor to reduce it to the combination of two smaller tensors, then by repeating this process $dN-1$ times we will arrive at the MPS representation of the tensor
\begin{equation}\label{eq:MPStensor}
     A_{\Pi_{i=1}^d \sigma_1^i, \dots, \sigma_N^i} = \sum_{\alpha_1}^{D_1} \sum_{\alpha_2}^{D_2} \dots \sum_{\alpha_{dN-1}}^{D_{dN-1}} A_{\alpha_1}^{\Tilde{\sigma}_1} A_{\alpha_1,\alpha_2}^{\Tilde{\sigma}_2} \dots A_{\alpha_{dN-1}}^{\Tilde{\sigma}_{dN}} \ ,
\end{equation}
where $\alpha_j$ denotes the singular values obtained from the SVD. Note that we have introduced the following to simplify the notation of Eq.~\eqref{eq:MPStensor}
\begin{align}
 \Tilde{\sigma}_n=\left\{\begin{array}{ll}
                  \sigma_n^1, & 1\le n\le N,\\[0.1cm]
                  \sigma_{n-N}^2, & N< n \le 2N,\\[0.1cm]
                  \mathrm{etc.}
                 \end{array}\right.
\end{align}
i.e., the first $N$ $\Tilde{\sigma}_n$ represent the first dimension, the consecutive $N$ then represents the second dimension, and so on. It should be noted that there are different ways to encode a high dimensional problem into MPS form \cite{Q_inspired_fluids,Plasma_MPS}.

\begin{figure}[t]
    \centering
    \includegraphics[width=1.0\linewidth]{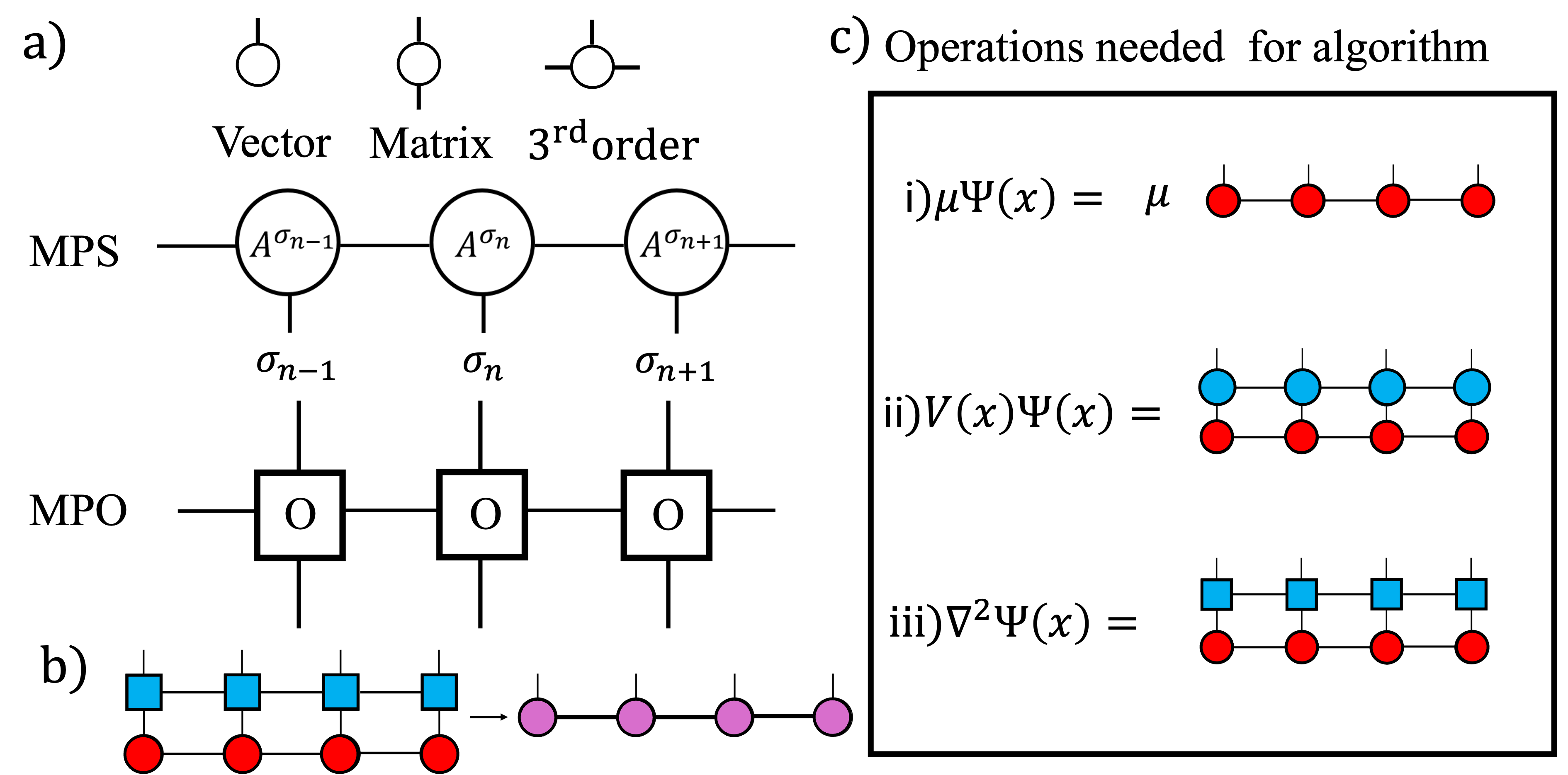}
    \caption{$\mathrm{a})$ Pictorial representation of tensor networks. Each tensor is represented as a circle or square, where the number of lines emerging from a tensor determines the tensor rank. Any two connecting lines between two objects represents a summation over that index (contraction), leading to the pictorial representation of an MPS and MPO as shown. $\mathrm{b})$ Pictorial representation of contracting an MPO and an MPS to produce a new MPS. $c)$ Operations on a MPS required for our tensor network simulation methods of the GPE, which consists of i) multuplying an MPS by a scalar constant, ii) multiplication of an MPS by a potential term, represented by a diagonal MPO, and applying differentiation operators (MPOs) onto an MPS. }
    \label{fig:MPS_pic}
\end{figure}

There exists a complementary pictorial representation of an MPS, illustrated in Fig.~\ref{fig:MPS_pic}a . Each $A^{\sigma_n}$ tensor is represented by a circle, with corresponding indices shown as lines emerging from the body. Any two connecting lines then indicate a summation over the shared index, known as a contraction, i.e. matrix multiplication. This diagrammatic representation will become useful when we come to discuss the full algorithm implemented.

\subsection{Truncation of a matrix product state}

The number of singular values in between each matrix is referred to as the bond dimension, and it directly gives the size of the corresponding dimension of the matrices on either side. The maximum bond dimension after performing the SVDs for Eq.~\eqref{eq:MPStensor} exactly is given by $D_n = min(2^n,2^{N-n})$. However, many of the singular values can be effectively zero and can be discarded allowing for a controlled truncation of the bond dimension without a significant loss in accuracy of the numerical approach. To do so, we perform an SVD at each bond within the MPS to generate singular values $S_i$, which are normalised such that $\sum_i S_i^2 =1 $. We then retain the $\chi$ largest singular values, whilst discarding the rest, thus performing a truncation of our state. Unless otherwise stated, we set a cutoff to determine $\chi$, where we discard any singular value which when squared is below this cutoff, along with their corresponding elements in the left and right matrices of the SVD. Using a cutoff for truncation is often advantageous as our bond dimension can dynamically increase or decrease during a simulation, and is useful in ensuring convergence of our results.

If an MPS provides a good representation of $\Psi$, i.e. we can represent $\Psi$ accurately with a small bond dimension, depends on the correlations in the data. 
By restricting the bond dimension in the MPS at a given bond $n$, we are effectively limiting the possible correlations between the two halves of the MPS on either side of bond $n$. Hence, the trick to the efficient encoding of a given state in an MPS is to find a representation such that correlations across all splits of the system at each given bond are minimal. We can then achieve physically motivated data compression through this truncation of the singular values. Additionally, by decomposing $\Psi$ into a tensor network we can infer information about the structure of correlations which are present in our state.

\subsection{Representing the GPE as a matrix product operator}

Just as one can encode classical functions into MPS form, one can express an operator in the form of a Matrix Product Operator (MPO) with a limited bond dimension. An MPO acts on an MPS to produce another MPS, and is thus represented pictorially as shown in Fig.~\ref{fig:MPS_pic}a. To apply an MPO to an MPS, one contracts with the MPS as shown in Fig.~\ref{fig:MPS_pic}b. Note that after the application of an MPO, the bond dimension of the resultant MPS will grow and the truncation of the bond dimension will be applied if applicable.

The application of the GPE given in Eq.~\eqref{eq:GPEmodified} requires the following operations on the MPS; (i) multiplication by a scalar, (ii) multiplication by a scalar function, (iii) the implementation of the kinetic term, (iv) squaring the MPS, and (v) the time-stepping procedure. These steps are illustrated in Fig.~\ref{fig:MPS_pic}c. The implementation of (i) is straightforward, with (ii) and (iv) being achieved by promoting the MPS to a diagonal MPO
\begin{equation}
    \Psi_\mathrm{MPO}= \sum_{\boldsymbol{\sigma}} A^{\Tilde{\sigma}_1}\delta^{\sigma_1}_{\sigma_1'} A^{\Tilde{\sigma}_2}\delta^{\sigma_2}_{\sigma_2'}  \dots A^{\Tilde{\sigma}_{dN}} \delta^{\sigma_{dN}}_{\sigma_{dN}'} \ket{\boldsymbol{\Tilde{\sigma}}}\bra{\boldsymbol{\Tilde{\sigma}'}}\ ,
\end{equation}
where $A^{\Tilde{\sigma}_n}$ are the MPS tensors and $\delta^{a}_{b}$ is the kronecker delta. Contraction of such a diagonal MPO onto an MPS results in performing an element wise multiplication of the encoded functions.

There are two main options for the implementation of the kinetic term in the GPE, more specifically, the implementation of the Laplace operator. The most direct approach is to encode the Laplace operator as an MPO \cite{QTT_ops}. This can be written out  by first defining effective raising and lowering operators 
\begin{equation}
    s_i^{-}=\begin{pmatrix}
        0 & 1 \\
        0 & 0
    \end{pmatrix} \ , \ s_i^{+}=\begin{pmatrix}
        0 & 0 \\
        1 & 0
    \end{pmatrix} , \ I_i=\begin{pmatrix}
        1 & 0 \\
        0 & 1
    \end{pmatrix},
\end{equation}
at the $i$th site of the MPS. We can then write the central finite difference in terms of these operators on the MPS as
\begin{widetext}
\begin{eqnarray}
    &\hat{\partial}_x^2 & =  \frac{\partial^2}{\partial x^2} = \frac{\hat{S}^+ +\hat{S}^- -2\hat{I}}{\Delta x^2} \\ \nonumber
    &=&\frac{1}{\Delta x^2} \begin{pmatrix}
        s^+_1 && s^-_1 && I_1
    \end{pmatrix} \begin{pmatrix}
        s^-_2 && 0 && 0\\
        0&&s^+_2 && 0\\
        s^+_2 &&s^-_2 && I_2
    \end{pmatrix} \begin{pmatrix}
        s^-_3&& 0 &&0\\
        0&&s^+_3 && 0\\
        s^+_3 &&s^-_3 && I_3
    \end{pmatrix} \cdots
     \begin{pmatrix}
        s^-_{N-1}&& 0 &&0\\
        0&&s^+_{N-1} && 0\\
        s^+_{N-1}&&s^-_{N-1} && I_{N-1}
    \end{pmatrix}
    \begin{pmatrix}
        s^-_N \\
        s^+_N \\
        s^+_N + s^-_N -2 I_N
        
    \end{pmatrix}.
\end{eqnarray}
\end{widetext}

One can identically encode higher-order finite-difference approximations, for example an $8$th order central finite difference approximation \cite{FD_stencils}. Another option is to utilise so-called split-step methods, a common approach for the simulation of the GPE, and utilise the quantum Fourier transform which we will discuss in the next section.

\section{The split-step method with tensor networks} \label{TN_QFT}

An alternative to the finite difference representations of $\nabla^2$ described above is to use the Fourier transform such that the differential terms become diagonal in momentum space. We can define the (discrete) Fourier transform $\boldsymbol{\mathcal{F}}$ as
\begin{equation}
   \tilde{f}(\mathbf{k})=\boldsymbol{\mathcal{F}}[f(\mathbf{r})] = \frac{1}{\sqrt{M}} \sum_{\mathbf{r}} e^{i \mathbf{k} \cdot \mathbf{r}} f(\mathbf{r}) \ ,
\end{equation}
where $M$ is the total number of data points.We can therefore write the differentiation operator along $x$ as
\begin{equation}
    \hat{\partial}_x =  -i \  \boldsymbol{\mathcal{F}^{-1}} \sum_{\mathbf{k}} k_x \ket{\mathbf{k}}\bra{\mathbf{k}} \boldsymbol{\mathcal{F}} \ ,
\end{equation}
with $k_x$ being the $x$-component of the, in general, three-dimensional quasimomentum $\mathbf{k} = \left(k_x,k_y,k_z\right)$. Then to obtain the kinetic term of the GPE, we can utilise the fact that
\begin{equation}
    \nabla^2 =  - \  \boldsymbol{\mathcal{F}^{-1}} \sum_{\mathbf{k}} \left(k_x^2 + k_y^2 + k_z^2\right) \ket{\mathbf{k}}\bra{\mathbf{k}} \boldsymbol{\mathcal{F}} \ .
\end{equation}

This approach of utilising the Fourier transform for the kinetic term forms the backbone of the split-step Fourier method to the simulation of the GPE \cite{javanainen2006symbolic,feit1982solution}. As the name suggests the split-step method breaks up the evolution into discrete steps in both time and between the kinetic and potential terms of the GPE. Here we will independently apply the kinetic and diagonal terms of the modified GPE of Eq.~\eqref{eq:GPEmodified} such that evolution for a time step $\Delta t$ under a second-order Trotter-Suzuki expansion would be given by
\begin{equation}
    U^{(2)}\left( \Delta t \right) = {\rm e}^{-i \Delta t H_d/2} \boldsymbol{\mathcal{F}}^{-1} {\rm e}^{-i \Delta t |\mathbf{k}|^2/2} \boldsymbol{\mathcal{F}} {\rm e}^{-i \Delta t H_d/2},
    \label{eq:2nd_split_step}
\end{equation}
with $|\mathbf{k}|^2 = k_x^2 + k_y^2 + k_z^2$ and $H_d = V(\mathbf{r}) -\mu + g|\Psi|^2$. One can additionally construct a fourth-order Trotter decomposition from the second-order \cite{Hatano_2005} to obtain
 \begin{equation}
      \hat{U}^{(4)}(\Delta t) =  \hat{U}^{(2)}(\delta_1)  \hat{U}^{(2)}(\delta_1)  \hat{U}^{(2)}(\delta_2)  \hat{U}^{(2)}(\delta_1)  \hat{U}^{(2)}(\delta_1) \ , \label{eq:4th_split_step}
 \end{equation}
where $\delta_1 = \frac{\Delta t}{4-\sqrt[3]{4}} $ and $\delta_2 = \Delta t -4\delta_1 $, which can provide a more accurate simulation, but at a cost of five times the number of operations per time-step. The error of the second order split-step implementation scales as $\mathcal{O}(\Delta t^2)$, whilst that of the fourth order implementation scales as $\mathcal{O}(\Delta t^4)$.

\subsection{The quantum Fourier transform}
The QFT is a well established operation that forms the backbone of many quantum algorithms such as Shor's algorithm \cite{nielsen2001quantum}. Given a quantum state $\ket{\Psi}=\sum_{\boldsymbol{\Tilde{\sigma}}}\Psi_{\boldsymbol{\Tilde{\sigma}}}\ket{\boldsymbol{\Tilde{\sigma}}}$,where $\ket{\boldsymbol{\Tilde{\sigma}}} = \ket{\Tilde{\sigma}_1,\Tilde{\sigma}_2,\Tilde{\sigma}_3,\cdots, \Tilde{\sigma}_{dN}}$, one can apply the QFT operator $\hat{\boldsymbol{\mathcal{F}}}_{\mathrm{QFT}}$

\begin{equation}\label{eq:MPSQFT}
\hat{\boldsymbol{\mathcal{F}}}_{\mathrm{QFT}} = \frac{1}{\sqrt{2^{dN}}}\sum_{\boldsymbol{\sigma},\boldsymbol{\sigma}'=0}^{2^{dN}-1} e^{\frac{2\pi i \boldsymbol{\Tilde{\sigma}} \boldsymbol{\Tilde{\sigma}}' }{2^{dN}}} \ket{\boldsymbol{\Tilde{\sigma}}}\bra{\boldsymbol{\Tilde{\sigma}}'} \ .
\end{equation}
In this way, the QFT is the quantum analogue of the discrete Fourier transform. One can construct a quantum circuit for the QFT, which consists of applying single qubit Hadamard gates ($H$), and a series of controlled rotation gates $R(\theta)$
\begin{equation}
    H=\frac{1}{\sqrt{2}}\begin{pmatrix}
        1 && 1 \\
        1 && -1
    \end{pmatrix} \ , \ R(\theta)= \begin{pmatrix}
        1 && 0 \\
        0 && e^{i \theta}
    \end{pmatrix} \ .
\end{equation}
To exactly implement the QFT as shown in Eq.~\eqref{eq:MPSQFT} one must additionally perform a final reversal of the ordering of the qubits. It has previously been shown that implementing this full operation, including the final qubit reversal, generates a large amount of entanglement, and the QFT corresponds to a maximally entangling operation \cite{Tyson2002OperatorSchmidt,QFT_schmidts2003}, which to implement as an MPO would require exponentially large bond dimensions.

However, it has recently been shown that if one instead implements the core part of the QFT operation and neglects the final reversal step, this then corresponds to an operator with exponentially decaying singular values, i.e. it generates little entanglement \cite{QFT}. 
This key observation allows the QFT to be efficiently implemented on classical computers as an MPO with small bond dimension $\chi_{\mathrm{MPO}}$.

\subsection{MPO construction}

\begin{figure}[htb!]
    \centering
    \includegraphics[width=.6\linewidth]{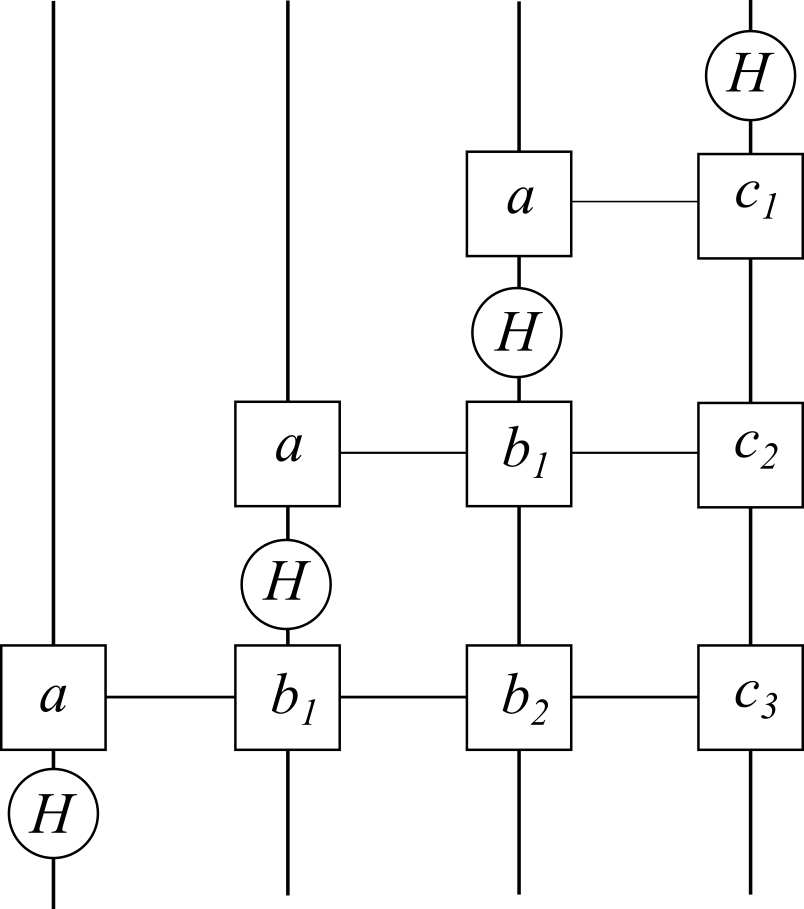} 
      \caption{Tensor network to implement the QFT on a 4 site MPS. Consists of applying local Hadamard gates $H$ onto each tensor, followed by the application of MPOs, constructed from $a$, $b_n$ and $c_n$ tensors given by Eq.~\eqref{eq:QFT_MPO_tensors}. One can contract the above network into a single MPO to implement the QFT. }
       \label{fig:QFT_TN}
\end{figure}

\begin{figure}[htb!]
    \centering
    \includegraphics[width=.7\linewidth]{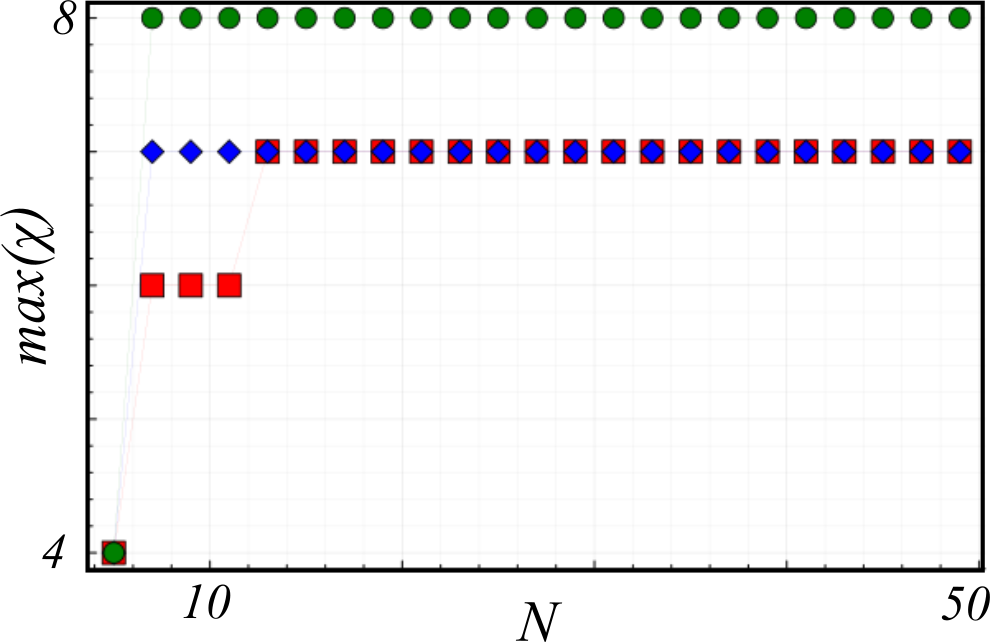}
    \caption{$\mathrm{a})$ Scaling of maximal bond dimension of the QFT MPO with increasing length of MPO $N$ (where the number of grid points in 1D scales as $M=2^N$). We show the maximal bond dimensions when constructing the QFT-MPO with a truncation cut-off of $10^{-10}$ (red square), $10^{-12}$ (blue diamond) and $10^{-15}$ (green dot) respectively. }
    \label{fig:QFT_plots}
\end{figure}

The QFT can be implemented as a tensor network as illustrated in Fig.~\ref{fig:QFT_TN}, which consists of contracting single tensor $H$ gates, along with a series of MPOs of bond dimension $2$. Each MPO implements a series of controlled rotation gates, and are constructed from three types of tensors $a,b_n$ and $c_n$ defined as
\begin{eqnarray}
   & a= \begin{pmatrix}
    \ket{0}\bra{0} && \ket{1}\bra{1}
\end{pmatrix}  \ , \ b_n = \begin{pmatrix}
    I && 0 \\
    0 && R(\frac{\pi}{2^n})
\end{pmatrix}  &\\ & c_n= \begin{pmatrix}
    I  \\
    R(\frac{\pi}{2^{n}}) 
\end{pmatrix}  \ . & \nonumber
\label{eq:QFT_MPO_tensors}
\end{eqnarray}
By contracting the tensor network, we obtain an overall MPO which performs the QFT operation. We analyse how the maximal bond dimension of this QFT-MPO in 1D ($d=1$) scales with increasing grid density, i.e. the length of the MPO ($N$). Note that an MPS or MPO of size $N$ encodes $2^{N}$ grid points as discussed previously. After contracting the network in Fig.~\ref{fig:QFT_TN}, we truncate the MPO back to a smaller bond dimension using a pre-defined truncation cut-off, with the results illustrated in Fig.~\ref{fig:QFT_plots}$a$ for various cut-offs. We observe the bond dimension of the QFT-MPO quickly saturates for low grid density to a small bond dimension, even in the case of a cut-off of $10^{-18}$, i.e., effectively numerically exact, with the maximum bond dimension saturating at $\chi_\mathrm{MPO}=10$.

It was shown in Ref.~\cite{QFT} that the entanglement properties of the QFT circuit are identical to those of time evolving a chain of qubits under a z-z interacting Hamiltonian, with exponential decaying interactions. Time evolution under such a Hamiltonian obeys a form of area law \cite{QFT,EntanglemntAreaLawsGong2017}, and for a chain of qubits, the entanglement generation becomes bounded to a constant, and does not increase with qubit number. This behaviour is thus exactly reflected in the saturating bond dimension of the QFT-MPO. 

\begin{figure}[t]
    \centering
    \includegraphics[width=\linewidth]{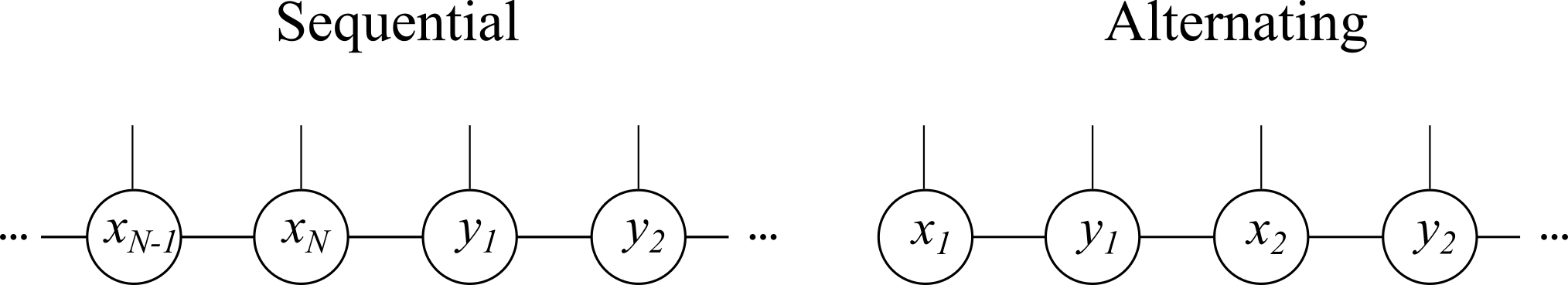}
    \caption{Two possible MPS orderings considering for $2D$ simulations of the GPE. $x$ and $y$ denote if the tensors encode features along the $x$ or $y$ direction respectively. By encoding all the $x$ tensors in the first half of the MPS followed by the $y$ tensors in the later half, we form a sequential encoding. Alternatively, we can group $x$ and $y$ tensors representing the same length scales in an alternating fashon, creating the Alternating encoding. }
    \label{fig:MPS_ordeings}
\end{figure}

Above we analysed the QFT-MPO for a one-dimensional problem. We now analyse the behaviour of the QFT-MPO for higher dimensions. We will analyse the scaling of the bond dimension of the QFT-MPO for the case of two different orderings, sequential and alternate ordering, as illustrated in Fig.~\ref{fig:MPS_ordeings}. Sequential ordering, as demonstrated in Sec.~\ref{Represeting as a TN}, uses the first $N$ tensors in the MPO to index the $x$ positions, and the remaining $N$ tensors for $y$. Alternate ordering groups together $x$ and $y$ tensors which encode the same length scale in an alternating pattern, i.e. $x_1 y_1 x_2 y_2 $. While we are free to choose any ordering, we will gain the most benefit from our MPS approach if we choose the ordering such that correlations between distant tensors of the MPS are minimised.

\begin{figure}[t]
    \centering
\includegraphics[width=\linewidth]{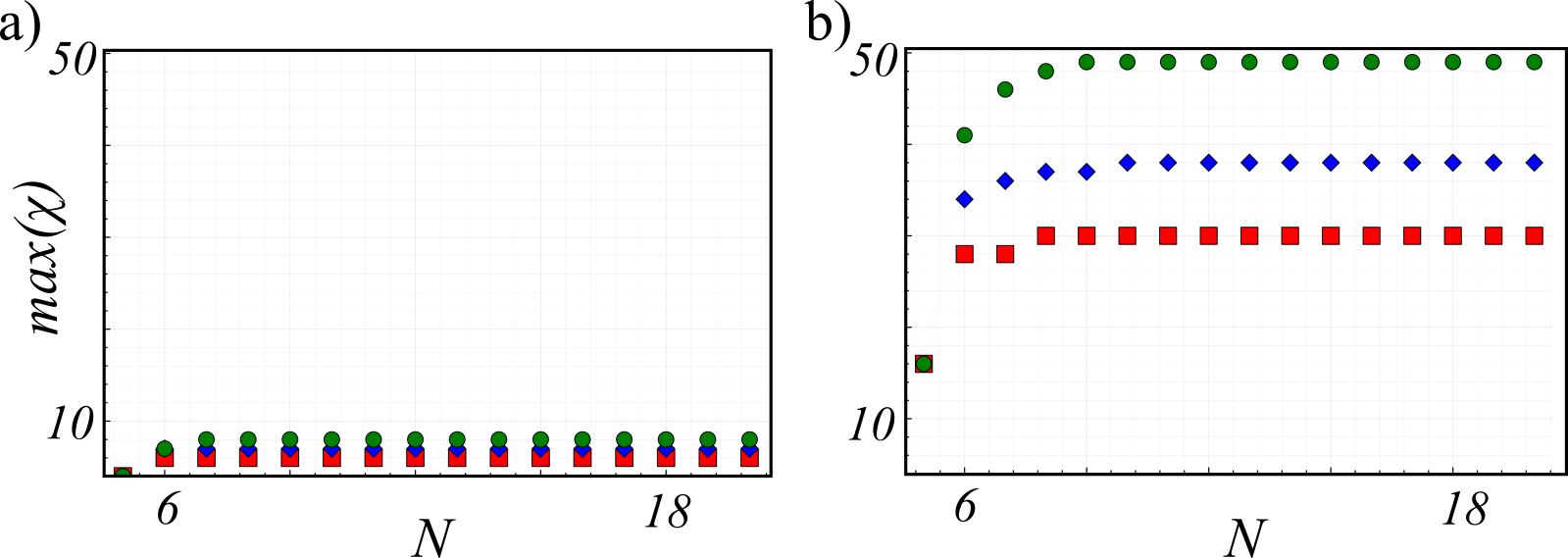}
    \caption{Maximal bond dimension of 2D QFT MPO for sequential ($\mathrm(a))$) and scale ($\mathrm(a))$) MPS orderings whilst utilising truncation cutoffs of  $10^{-12}$ (red squares), $10^{-15}$ (blue diamonds) and $10^{-18}$ (green dots). We plot the maximal bond dimensions as a function of increasing MPO length $2N$, which represents a $2^{N}\times 2^{N}$ spatial grid. }
    \label{fig:2D_QFT_bdims}
\end{figure}

The two-dimensional QFT-MPO was obtained by first constructing the QFT operator for $x$ and $y$  ($\mathcal{F}_x, \mathcal{F}_y$) separately, and then contracting these two MPOs to form the full two-dimensional operator.
The maximal bond dimensions of the 2D QFT show an identical scaling to the 1D case for both MPO orderings, with a bond dimension which quickly saturates and doesn't scale with MPO length, as shown in Fig.~\ref{fig:2D_QFT_bdims} for different truncation cutoffs and both MPS orderings. The choice of ordering, however, does have a significant impact on the resultant bond dimension. The Sequential ordering reaches a maximal bond dimension exactly equal to the 1D case of $\chi_\mathrm{MPO}=10$ for a truncation cutoff of $10^{-18}$. However, the Alternate ordering saturates at $\chi_\mathrm{MPO}=65$,  more than double compared to the Sequential order. To construct the two-dimensional QFT MPO, we apply $\mathcal{F}_x$ and $\mathcal{F}_y$ separately. The $\mathcal{F}_x$ operator only generates entanglement between $x$ tensors, and likewise for the $\mathcal{F}_y$ operation. For the sequential encoding, $x$ and $y$ are separated, and thus we are in effect simply applying a 1D QFT on the first $N$ tensors, and then a QFT on the remaining $N$ tensors independently.  In contrast, for the alternate ordering, correlations between $x$ tensors have to be transferred through the MPO via intermediary $y$ tensors, and vice-versa. Thus to perform the two-dimensional QFT, each bond in the MPO now needs to capture the generated correlations between the $x$ and $y$ tensors, i.e. effectively double the entanglement.
As the entanglement entropy $S$ generally scales as $S\sim \log(\chi)$, the required bond dimension for the alternate ordering should be much more than double the $1$D QFT bond dimension, as is observed. Due to the significant reduction in bond dimension for the Sequential ordering, we use this ordering throughout unless otherwise stated.

We have illustrated how the Fourier transform can be implemented as an MPO of small bond dimension, and as such, we can perform a fourier transform by contracting with an MPS at a computation cost scaling as $\mathcal{O}(N\chi_\mathrm{MPS}^2) = \mathcal{O}(\log_2(M)\chi_\mathrm{MPS}^2)$, where $M$ is the number of gridpoints in the simulation. This contrasts with the computation scaling of the standard FFT, which scales as $\mathcal{O}(M\log(M))$, and points to the fact that performing the fourier transform is no longer the most costly operation when exploiting MPS. Instead, calculation of the non-linear term $|\Psi|^2$ dominates the computation, scaling as $\mathcal{O}(N\chi_\mathrm{MPS}^4)$.

\subsection{Analytical example}

\begin{figure}[t!]
    \centering
    \includegraphics[width=\linewidth]{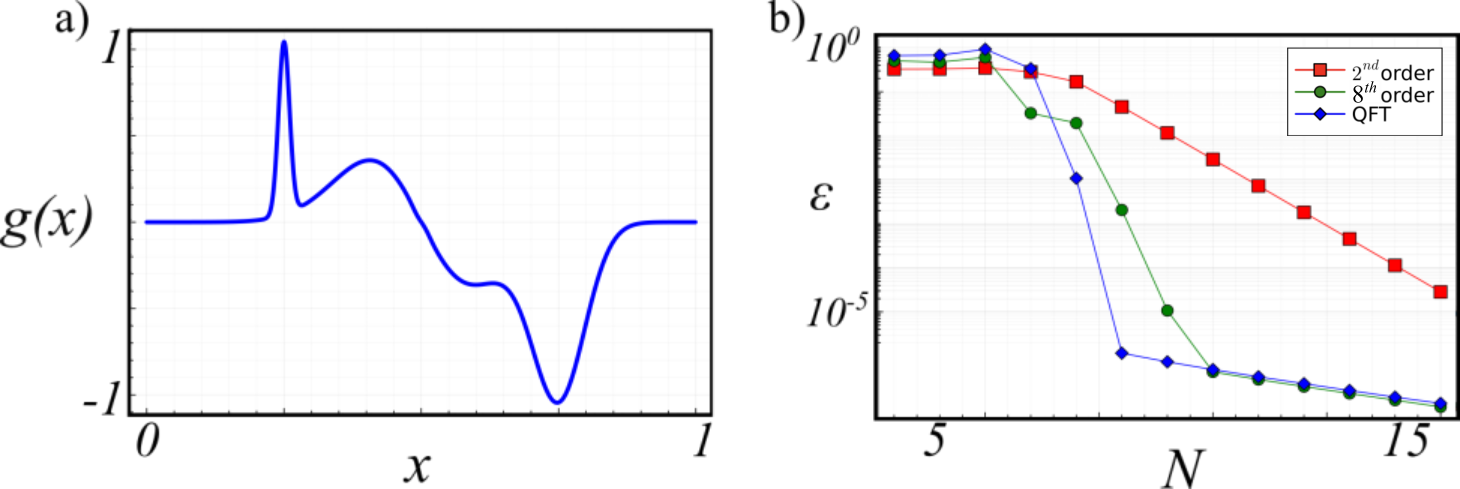}
    \caption{$\mathrm{a})$ Test function $g(x)$ where $x=[0,1]$ , as defined in Eq.~\eqref{eq:test_func}, to test our different MPO differentiation techniques.  b) Error between the exact analytic derivative of our test function $g(x)$ and the derivative calculated via a $2nd$ order central finite difference MPO (red square), an $8$th order central finite difference MPO (green dots) and the QFT based MPO derivative (blue diamonds), plotted as function of increasing MPO length $N$. }
    \label{fig:der_example}
\end{figure}

We consider a simple test case to demonstrate the ability of the QFT-MPO to perform differentiation on a chosen test function $g(x)$
\begin{widetext}
\begin{equation}
    g(x) = \sin{(2 \pi x)} \left[ e^{\frac{-(x-a)^2}{\delta_1}} +  e^{\frac{-(x-\frac{a}{2})^2}{\delta_2}} - \frac{e^{\frac{-(x-a)^2}{\delta_2}}}{4} +   e^{\frac{-(x-\frac{3a}{4})^2}{\delta_3}} \right] \ . \label{eq:test_func}
\end{equation}
\end{widetext}
We will consider the case of periodic boundary conditions, which allows for a simpler implementation of the Fourier transform without the need to truncate low quasimomentum due to a boundary. We plot this function in $x \in [0,1]$ in Fig.~\ref{fig:der_example}a, with $\delta_1=0.1$,  $\delta_2 =0.01 $, $\delta_3=0.05$ and $a=0.5$.

For a direct comparison of the QFT and finite difference methods outlined previously, we will calculate the error $\varepsilon$ for different methods as
\begin{equation}
    \varepsilon_\mathrm{MPO} = \frac{1}{2^N} \sqrt{\sum_x \left[g'_\mathrm{MPO}(x) - g'_\mathrm{Exact}(x)\right]^2 } \ 
    \ ,
\end{equation}
where $g'_\mathrm{Exact}(x)$ is the exact analytical derivative, 
and $g'_\mathrm{MPO}(x)$ is the derivative obtained from a given implementation as an MPO. In Fig.~\ref{fig:der_example}b, we show $\varepsilon_\mathrm{MPO}$ using the QFT tensor network approach, a $2nd$ order central finite difference scheme, and an $8$th order finite difference stencil \cite{Q_inspired_fluids} for an increasingly fine spatial resolution. The QFT tensor network approach shows small errors for dense grids similar to the $8$th order finite difference and provides a much smaller error than the $2nd$ order finite difference. For all numerical differentiation schemes, we see that our errors tend to machine precision, i.e. exact agreement with the analytic result, as one increases the number of grid points.

\section{Examples of simulating the Gross-Pitaevskii equation with tensor networks}

We will now consider a number of examples of problems commonly studied utilising the simulation of the GPE. We will utilise the tensor network approach introduced above using both finite difference and split-step methods for the application of the kinetic term. When utilising the finite-difference approach with tensor networks we use an $8$th order finite difference stencil, and we perform time-evolution using a 4th order Runge-Kutta scheme (RK4) \cite{Press2007}. RK4 schemes are widely used in simulations of differential equations \cite{Press2007}. To perform one time-step, one first calculates $\frac{\partial \Psi(t)}{\partial t}$, which is then added to $\Psi(t)$ to produce an intermediate increment of the solution. Four intermediate increments are calculated, which are then averaged over to obtain the next time-step, $\Psi(t+\Delta t)$ . Further details are provided in Appendix \ref{App:RK4}. The split-step approach by contrast utilises the QFT and splits apart the evolution into real space and momentum space, as outlined in Eq.~\eqref{eq:2nd_split_step} and \eqref{eq:4th_split_step}.  We compare using both second and fourth order split-step methods.

\subsection{Soliton propagation}

\begin{figure}[t]
    \centering
    \includegraphics[width=.8\linewidth]{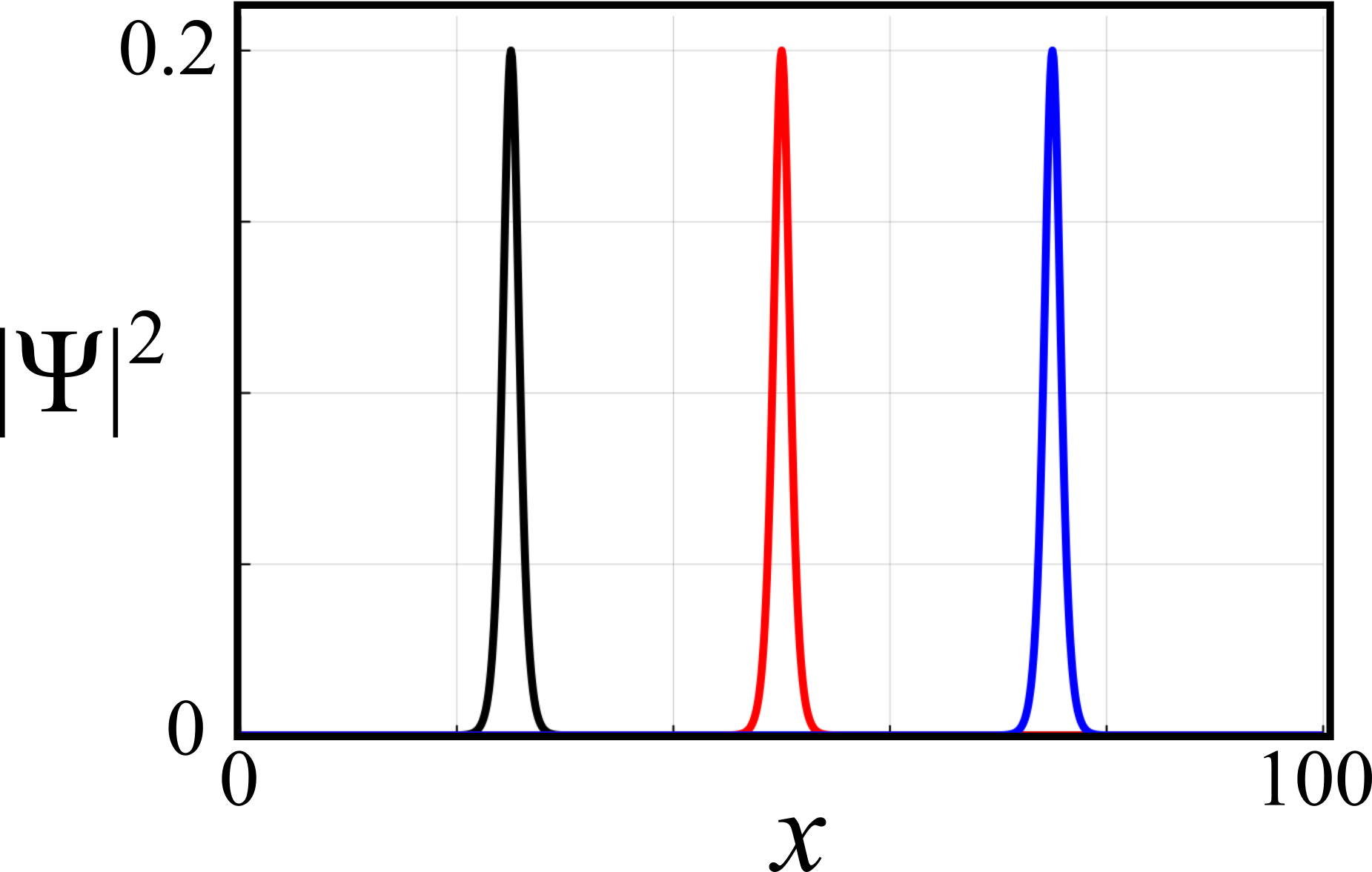}
    \caption{Soliton propagation over a spatial domain $x\in [0,100]$. A bright soliton is initialised at $t=0$ (black, leftmost pulse) and is left to propagate under the GPE with interaction strength $g=-5$ and a velocity of $v=1$ and amplitude $A=\frac{1}{\sqrt{|g|}}$ as prescribed in Eq.~\eqref{eq:soliton}. We plot snapshots of $|\Psi|^2$ at times of $t=10$ (red, middle pulse) and $t=20$ (blue, rightmost pulse) }
    \label{fig:Soliton}
\end{figure}

\begin{figure}[t]
    \centering
    \includegraphics[width=\linewidth]{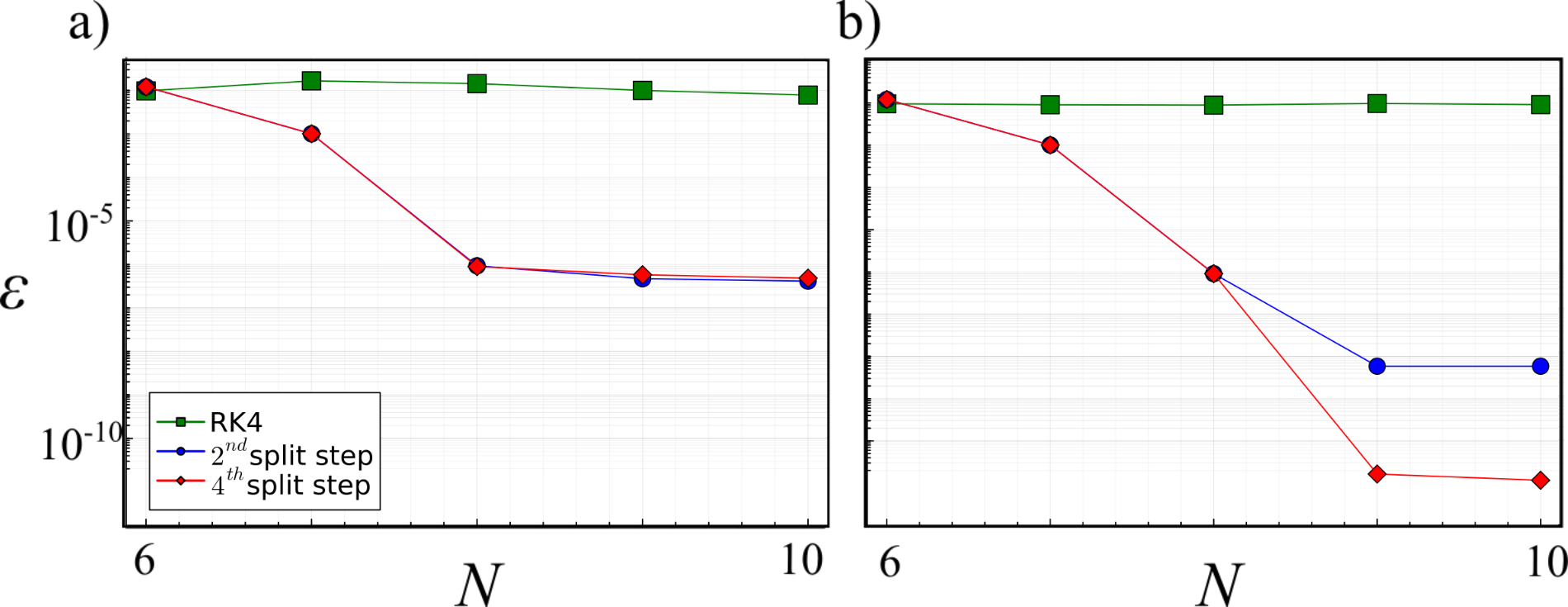}
    \caption{Errors relative to analytic solution for bright soliton propigation, with parameters outlined in Fig.~\ref{fig:Soliton}, from tensor network simulations using a finite difference RK4 scheme (green squares), a second order split-step scheme (blue dots) and a fourth order split-step evolution scheme (red diamonds) . We consider using an MPS truncation cutoff of $\mathrm{a})$ $10^{-12}$ and $\mathrm{b})$ $10^{-28}$ (set to ensure we retain all non-zero singular values, i.e. untruncated) across a range of MPS lengths ($N$), corresponding to $2^N$ grid points. Results are calculated with a time step of $\Delta t=0.001$, and compared at simulation time of $t=0.5$.  }
    \label{fig:Soliton_errs}
\end{figure}

As a first demonstration of performing time evolution, we consider the dynamics of a soliton. A soliton is an exact solution of the GPE, and consists of a localised wave packet in an otherwise uniform density ($|\Psi|^2$), which travels without dissipation, $\gamma=0$ (see Fig.~\ref{fig:Soliton}) \cite{Zakharov1970ExactTO,Soliton_Salasnich_2017}. We initialise the soliton at $t=0$, and use the different MPS simulation methods on offer to compare against the exact analytical solution for the soliton problem at later times. 

We will consider the case of a bright soliton, which is a solution to the GPE for attractive interactions, $g<0$. The analytical form the soliton takes on is given as
\begin{equation}
    \Psi(x,t) = A \: \mathrm{sech}\left(B (x-vt) \right) e^{i(vx -\frac{v^2t}{2}-\mu t)} \ ,
    \label{eq:soliton}
\end{equation}
where $A=\frac{B}{\sqrt{|g|}}$, $\mu = - \frac{B^2}{2}$, and $v$ is the velocity of the soliton in motion. For the simulations presented, we consider $g=-5$, $A=\frac{1}{\sqrt{|g|}}$ and $v=1$. We initialise $\Psi(x,t=0)$ in MPS form, and then use the above discussed time-evolution techniques to model the propagation of the soliton.
We quantify the accuracy of each MPS simulation method by defining the error $\varepsilon_\mathrm{t-e}$ as
\begin{equation}
        \varepsilon_\mathrm{t-e}= \sqrt{\frac{1}{2^N} \sum_j |\Psi^\mathrm{sim}_j-\Psi^\mathrm{Exact}_j|^2} \ , 
\end{equation}
where $\Psi^\mathrm{sim}$ is the results from our tensor network simulations, and $\Psi^\mathrm{Exact}$ is the analytical expression for the soliton.

We first explore the decay of errors as one increases the MPS length $N$, corresponding to resolving finer and finer grids, for the various simulation methods, shown in Fig.~\ref{fig:Soliton_errs}. We use both split-step methods and the finite difference RK4 scheme with a time-step of $\Delta t= 0.001$. We observe that across all resolutions, the split-step methods appear always to produce errors smaller than those from the RK4 scheme. 

We consider a truncation cutoff of $10^{-15}$ in Fig.~\ref{fig:Soliton_errs}a, and perform no truncation in Fig.~\ref{fig:Soliton_errs}b i.e. retain all non-zero singular values. In the case of the larger truncation cutoff, both the second and fourth order split-steps schemes produce very similar errors, and there is no clear benefit from the fourth order scheme. It is only when one reduces the truncation cutoff further that one observers the benefit from the fourth order split-step scheme. This is alsolinked to the contribution of the time-stepping to the errors, as we will now discuss. 

\begin{figure}[t]
    \centering
\includegraphics[width=\linewidth]{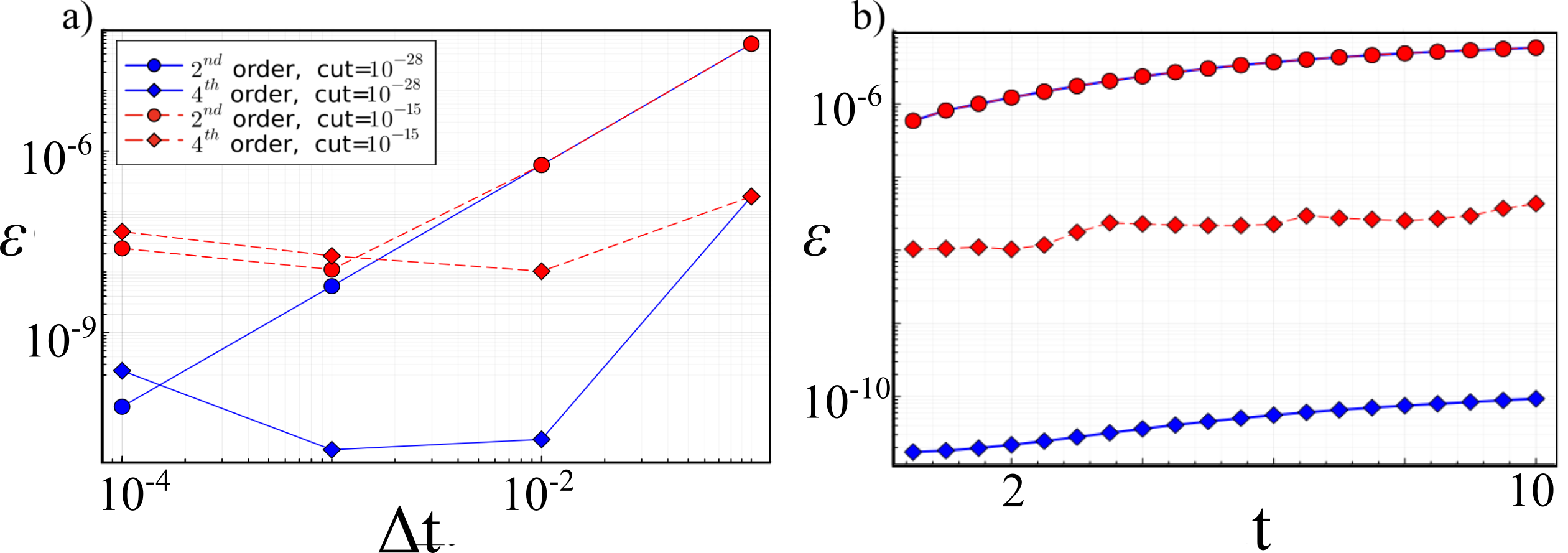}
    \caption{Comparison of second (dots) and fourth (diamonds) order split-step methods for soliton propagation test case, using truncation cutoffs of $10^{-28}$ (solid blue) and $10^{-15}$ (dashed red). $\mathrm{a})$ Error relative to analytic solution as a function of time-step $\Delta t$ at a simulation time of $t=0.5$ for a spatial grid consisting of $1024$ points. $\mathrm{b})$ Errors relative to analytic solution as a function of time as the simulation progresses on an $1024$ spatial grid with a time-step of $\Delta t=0.01$. }
    \label{fig:Soliton_SS_comparison}
\end{figure}

Increasing the spatial resolution of a simulation method alone will not necessarily reduce the errors to arbitrarily small values. One must also consider the time-step $\Delta t$. To this effect, we consider a fixed MPS length of $N=10$ and perform the above soliton simulation with a range of time-steps for the second order split-step methods, shown in Fig.~\ref{fig:Soliton_SS_comparison}$\mathrm{a})$. For a truncation cutoff of $10^{-28}$, as one decreases the time-step from $\Delta t=0.1$ to $\Delta t=10^{-4}$, the errors of the simulation decrease. If one instead uses a truncation cutoff of $10^{-15}$, the errors initialy decay with decreasing time-step, before begining to increase as one decreases the time-step below $\Delta t=10^{-3}$.

In the case of performing MPS contractions exactly with zero truncation, as one decreases time-step, the global simulation error should in general decrease for a sufficiently large spatial grid, as it does when dealing with a direct numerical simulation. However, performing truncation after each time step causes this logic to fail. More time-steps means performing more truncations to evolve to the same physical time in the simulation, and hence we end up discarding more information about the state and thus observe an increase in errors as one reduces the time-step. Thus, one needs to balance between time-step and truncation errors for a desired amount of data compression.

In addition, we compare how using the fourth order split-step method compares against its second order counterpart, also plotted in Fig.~\ref{fig:Soliton_SS_comparison}. Using the fourth order method, we notice a much stronger timestep-truncation interplay. This can be understood as to perform one time-step, the fourth order method performs 5 times the number of operations as the second order split-step approach. After each MPS operation, one performs a truncation of the bond dimension up to a set truncation cutoff, and thus to evolve one time step, we perform more truncations with the fourth order method, introducing larger truncation errors per time step and off-setting the benefit of the higher order method.

However, the power of the fourth order split-step method is that it allows one to choose a much larger time-step whilst retaining a high accuracy, as can be demonstrated in both Fig.~\ref{fig:Soliton_SS_comparison}a and Fig.~\ref{fig:Soliton_SS_comparison}b. For a time-step of $\Delta t=0.1$, using a second order split-step approach produces errors on the order of $10^{-5}$. Instead, using the fourth order scheme, we are able to achiever errors below the $10^{-10}$ level.

These results highlight the importance of balancing the time-step and desired truncation level correctly. For sufficiently small time-steps, the overall error of our tensor network simulations is due to an interplay between truncation level, time discretisation error and spatial discretisation error, which must all be balanced to obtain the most amount of data compression whilst minimising simulation errors.

\begin{figure*}[t]
          \centering
    \includegraphics[width=\linewidth]{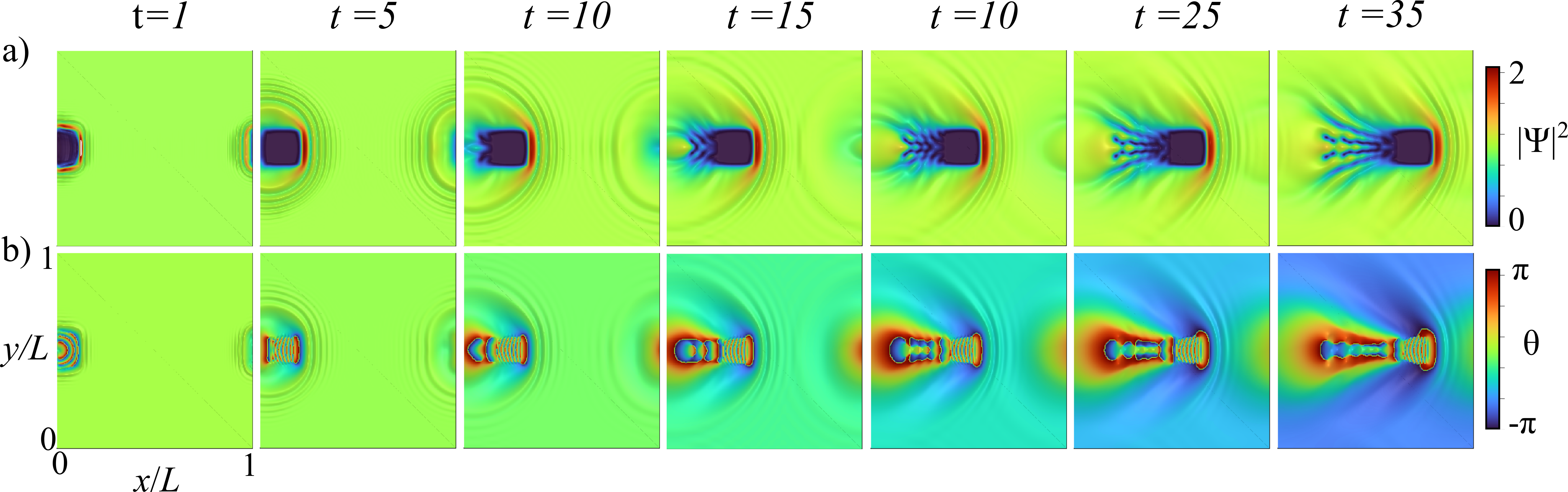}
    \caption{Demonstration of vortex shedding dynamic. A strong rectangular paddle potential is imposed on the BEC at $t=0$ centred on $x=0$, and $y=L/2$. This paddle potential is then swept across the BEC with a fixed sweep velocity $v_{sweep}=1.5$, which results in the formation of vortices behind the paddle as it moves through the system. We display the condensate density $|\Psi^2|$in a) , and the phase profile in b) at different snapshots in time $t$ as the paddle sweeps across. 
    These simulations were conducted using MPS split step method using a truncation cutoff of $10^{-12}$ on a $2^8 \times 2^8$ spatial grid of size $L=64$, with a time-step of $\Delta t=0.001$, interaction term $g=1$ and fixed dissipation parameter $\gamma=0.1$.  }
    \label{fig:paddles_sweep}
\end{figure*}

\subsection{Vortex shedding} \label{Sec:GPE_shed}

We now consider a more challenging dynamical example of vortex generation, which does not have an exact analytical solution. There has been a range of experiments conducted in recent years studying the behaviour and dynamics of vortices generated within superfluids \cite{Criticalvelocityforvortexshedding,Superfluid_Flow_BEC_PhysRevLett,Critical_Velocity_PhysRevLett,Vortex_Dipoles_PhysRevLett,Kwon_2016}, and understanding the behaviour and dynamics of vortices is central to the field of quantum turbulence. For example, it has been shown that the ability of a superfluid to flow without viscous effects is broken above a critical velocity due to the formation of vortices, which induce an effective viscous force \cite{super_Re}. Vortices in a BEC can be a localised region with zero occupancy in the condensate density, i.e. where $|\Psi|^2=0$. In addition, vortices formed in a BEC must have a quantised circulation, where the phase $\theta$, defined as $\Psi=|\Psi|e^{i\theta}$, must change by an integer multiple of $2\pi$ around the vortex core \cite{PhysRevLett.83.2498}. This quantisation arises from the fact that $\Psi$ must be single valued and continuous.
We will utilise the modified GPE with a non-zero dissipation term ($\gamma \neq 0$), as most studies into the generation of vortices include damping to account for atom loss. 

\subsubsection{Single Paddle}
\label{sec:single_paddle}

We use our MPS based methods to now study the dynamics of a BEC when a strong external potential is swept through the system. This external potential, $V_{\mathrm{pad}}(x,t)$, which we call a `paddle', has a fixed, rectangular shape, and is moved across the BEC with a sweep velocity $v$ (see Fig.~\ref{fig:paddles_sweep}). To have a resemblance closer to that which occurs experimentally, we implement a form of rectangular paddle in MPS form with rounded edges (see App. \ref{App:Rounded_paddles} for further details). Depending on the sweep velocity and the strength of dissipation, pairs of vortices will be generated from the rear of the paddle at regular intervals, which then trail behind the paddle as it moves across the system, akin to the formation of von-Karman vortices in classical fluid flows \cite{Kwon_2016}.  These vortices can be seen in Fig.~\ref{fig:paddles_sweep}a. We additionally display the phase ($\theta$) in Fig.~\ref{fig:paddles_sweep}b. Such phase plots allow vortices to be clearly identified, as the phase will change by an integer multiple of $2 \pi$ around the vortex. One can also infer the direction of rotation of a vortex, depending on if the phase changes from $-\pi$ to $\pi$ in a clockwise or counter clockwise direction.

\begin{figure}[t]
          \centering
    \includegraphics[width=\linewidth]{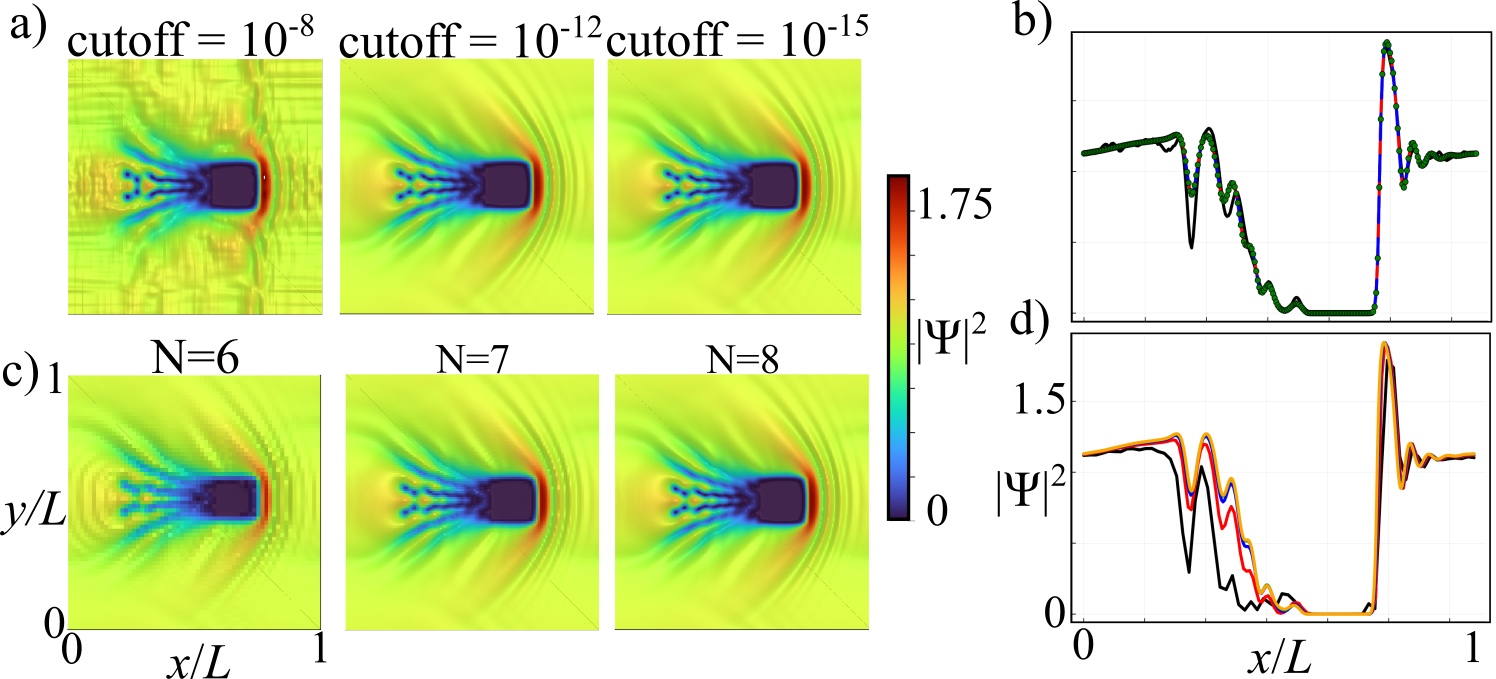}
    \caption{Convergence plots for vortex formation during the vortex shedding simulations. We compare using different spatial grid densities (MPS lengths) and truncation cutoffs and compare the resultant solutions of  $|\Psi|^2$ for convergence. We study convergence for identical simulation parameters as in Fig.~\ref{fig:paddles_sweep} at a simulation time of $t=28$ $a)$ Simulation on an $2^8\times2^8$ spatial grid for three truncation cutoffs of $10^{-8}, 10^{-12}$ and $10^{-15}$. We additionally plot a cut along the $y=\frac{L}{2}$ line in fig $b)$ for cutoff of $10^{-8}$ (solid black) , $10^{-10}$ (solid red), $10^{-12}$ (dashed blue), and $10^{-15}$ (green dots) . $c)$ Simulations are run using a consistent truncation cutoff of $10^{-12}$ but for increasing MPS lengths, i.e spatial resolution. Again, a cut along the $y=\frac{L}{2}$ line is displayed in $d)$, for MPS lengths of $N=6$ (black) , $7$ (red), $8$ (blue), $9$ (green) and $10$ (gold). }
    \label{fig:paddles_converge}
\end{figure}

As with all MPS methods, one must choose a suitably small truncation cutoff to achieve convergence of the physics, however we must also take into consideration the effect of the spatial resolution. We must ensure that the simulation contains sufficiently many grid-points to fully resolve vortex formation. We perform a vortex shedding simulation using the second order split-step method for a range of truncations and system sizes, and plot the resultant density profiles in Fig.~\ref{fig:paddles_converge}.  We observe that a truncation cutoff of $10^{-12}$ is more than sufficient to accurately converge the BEC dynamics, agreeing perfectly with truncation cutoffs of $10^{-15}$ across all simulation times and system sizes considered. In addition, we find that we really require a spatial density above $2^8\times 2^8$ to perfectly converge the physics. Performing simulations at a spatial density of $2^7 \times 2^7$ yields results which are qualitatively similar to the more resolved simulations, but there are noticeable deviations of the density around the $10\%$ level where the vortices form. 

\begin{figure}[t]
    \centering
    \includegraphics[width=\linewidth]{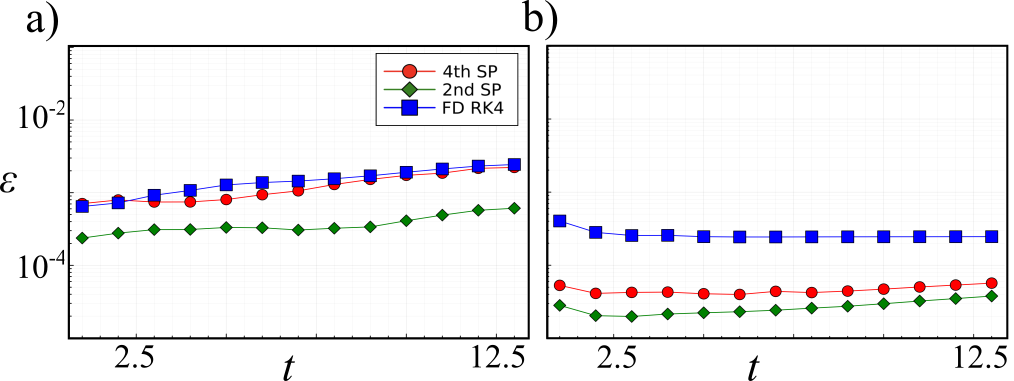}
    \caption{$\mathrm{a})$ Errors from MPS based vortex shedding simulations against XMDS2 direct simulation, for an $256\times 256$ spatial grid, with $L=64,g=1,\gamma=0.1, v=1.5$ and $\Delta t=0.001$. We compare errors during time evolution using the finite difference RK4 evolution scheme (blue squares), a second order split-step scheme (green diamonds) and a fourth order split-step scheme (red dots). The MPS simulations are run for different truncation cutoffs of $\mathrm{a})$ $10^{-12}$ and $\mathrm{b})$ $10^{-15}$.}
    \label{fig:dt=0.001_shed_errs}
\end{figure}

Having implemented the vortex shedding dynamics with MPS methods, we now wish to quantify the errors in these simulations, comparing the error from using MPS with a finite differences RK4 scheme, and from MPS exploiting the split-step methods using the QFT-MPO. To quantify the error of our MPS based simulations, we perform a comparison with direct numerical simulations, for which we use the software package of XMDS2 \cite{XMDS2DENNIS2013201}. XMDS2 solves ordinary and partial differential equations via a variety of direct numerical simulation methods. For the results presented in this work, we setup XMDS2 such that it is $4th$ order accurate in time, and calculates our spatial derivates in a spectral fashion, making use of the fast fourier transform. The computational costs of these methods scale as $M \log M$ with the number of grid points $M$. For comparison, within our MPS we are able to represent a number of grid points that grows exponentially with the length of our MPS. We define the global error from the MPS simulation as
\begin{equation}
\varepsilon_\mathrm{t-d}=\sqrt{\frac{1}{4^{N}} \sum_x |\Psi_\mathrm{MPS}(x)-\Psi_\mathrm{DNS}(x)|^2} \ ,
\end{equation}
where $\Psi_\mathrm{DNS}$ is the simulation result from XMDS2.

We first conduct a simulation on an $256 \times 256$ spatial grid, with a sweep velocity of $v=1.5$, and a dissipation rate of $\gamma=0.1$. We calculate the global error from different simulation methods in Fig.~\ref{fig:dt=0.001_shed_errs}. This simulation is conducted using a time-step of $\Delta t=0.001$ , and we observe that at all truncation errors, both the second-order and fourth-order split-step methods outperform the finite-difference RK4 method, as these produce smaller global errors as in Fig.~\ref{fig:dt=0.001_shed_errs}.

\begin{figure}[t]
    \centering
    \includegraphics[width=.7\linewidth]{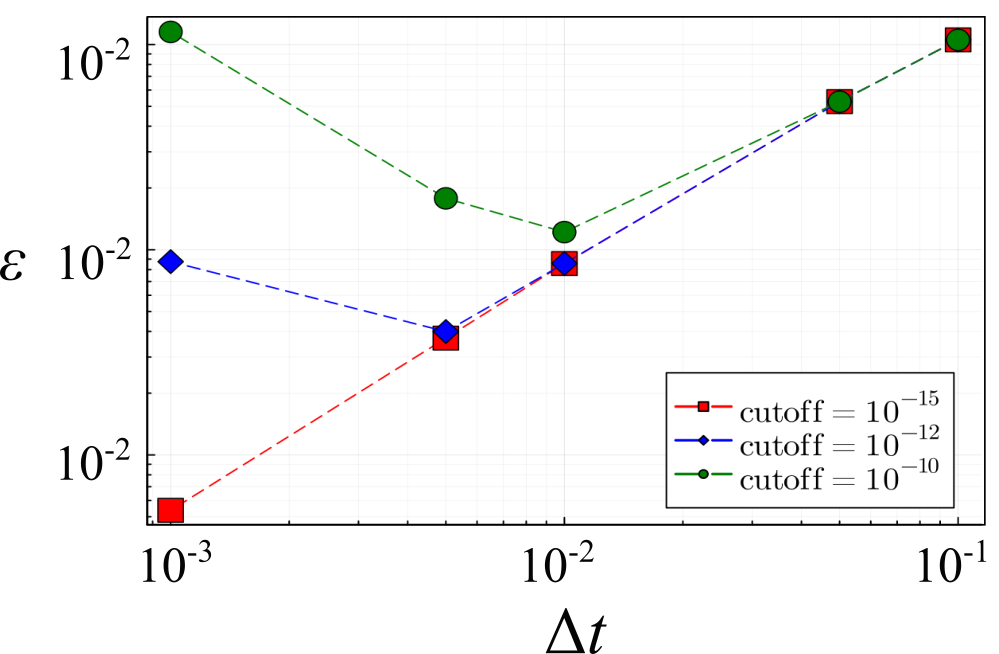}
    \caption{Vortex shedding simulation errors agaisnt XMDS2 as a function of time step ($\Delta t$). These results are from a second order split-step method, on an $128 \times 128$ spatial grid with $L=64,g=1,\gamma=0.1$ and sweep velocity $v=1,5$. Errors are plotted for truncation cutoffs of $10^{-10}$ (green dots), $10^{-12}$ (blue diamonds) and $10^{-15}$ (red squares). }
    \label{fig:dt_truncation}
\end{figure}

When performing the tensor network simulations with a truncation cutoff of $10^{-15}$, the fourth order split-step method produces the smallest errors on the order of $5\times10^{-7}$, which remains a constant error across the simulation time-frame. For the same truncation level, the second order split-step also performs well, but with errors which tend to increase to over $10^{-6}$ by the end of the simulation. However an important note is that the second-order accurate split-step method seems to produce smaller errors than the fourth-order split-step for larger truncation cutoffs of $10^{-12}$ and $10^{-10}$. This observation is also linked to the effect of time-steps on the simulation errors, as illustrated in Fig.~\ref{fig:dt_truncation}. The cause of such a phenomena was addressed in the previous soliton example.

\begin{figure}[t]
    \centering
    \includegraphics[width=\linewidth]{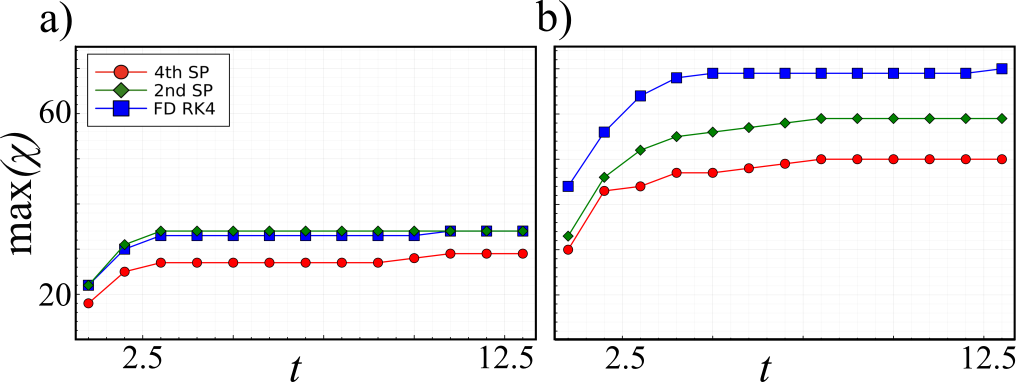}
    \caption{$\mathrm{a})$ Maximal bond dimension of $\Psi$ during vortex shedding simulations for an $128\times 128$ spatial grid, with $L=64,g=1, \gamma=1.0, v=1.5$ and $\Delta t=0.001$. We plot the maximal bond dimension during time evolution using the finite difference RK4 evolution scheme (blue squares), a second order split-step scheme (green diamonds) and a fourth order split-step scheme (red dots). The MPS simulations are run for different truncation cutoffs of $\mathrm{a})$ $10^{-12}$ and $\mathrm{b})$ $10^{-15}$. }
    \label{fig:dt=0.001_shed_Bdims}
\end{figure}

\begin{figure}[t]
    \centering
    \includegraphics[width=\linewidth]{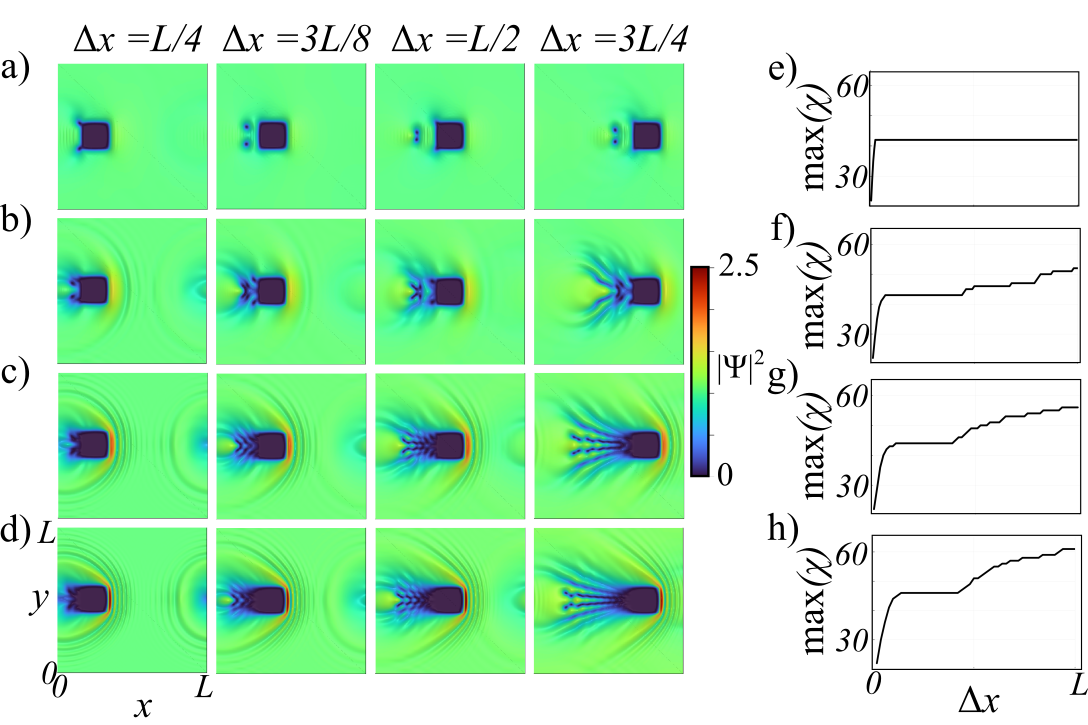}
    \caption{Snapshots of density $|\Psi|^2$ for paddles sweeping through a BEC with velocities $a) v=0.5$ and $b)v=1.0$, $c)v= 1.5$ and $d) v=2.0$. These simulations were conducted on a $256\times256$ spatial grid, using a truncation cutoff of $10^{-12}$. The corresponding bond dimensions are displayed in fig $e-h)$ showing the maximal bond dimensions of the MPS during the simulation as a function of how far the paddles have swept through the condensate $\Delta x$. For these simulations, we use $L=64,g=1,\gamma=0.1$ and $\Delta t=0.01 $.}
    \label{fig:vel_snaps}
\end{figure}

\begin{figure}[t]
    \centering
    \includegraphics[width=\linewidth]{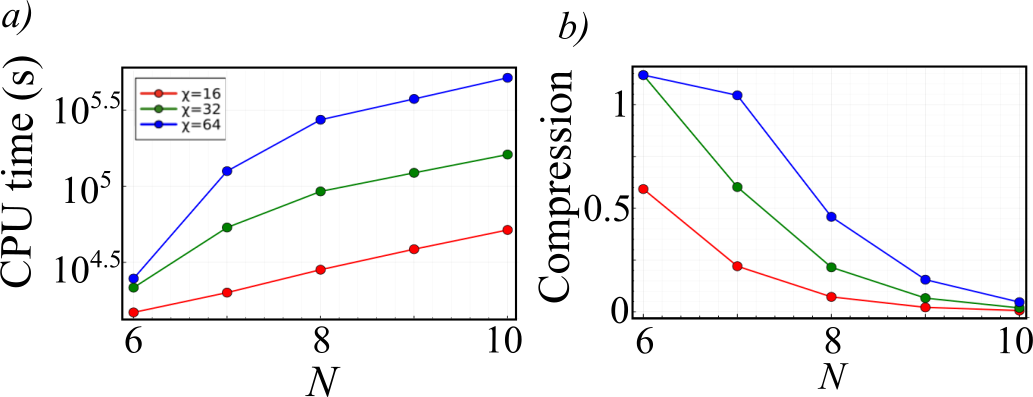}
    \caption{We perform instead fixed bond dimension simulations of vortex shedding  with $\Delta t=0.0005,\gamma=0.1,v=1.5$ and $L=64$ and analyse the behaviour of the simulation at time $t=5$. $\mathrm{a})$ Total CPU time to perform simulations up to $t=5$ on increasing grid resolutions ($N$) for  bond dimensions $\chi = 16$ (red), $32$ (green) and $64$ (blue). These codes were run on a 2021 M1 mac-book air. $\mathrm{b})$ Amount of compression we achieve by representing the solution at $t=5$ in MPS format with bond dimensions of $\chi=16,32 \mathrm{and} 64$ on increasing grid resolutions. We define the ratio of number of terms within MPS the representation to number of terms required during direct numerical simulation as Compression.
    }
    \label{fig:shed_bench}
\end{figure}

\begin{figure}[t]
    \centering
    \includegraphics[width=.9\linewidth]{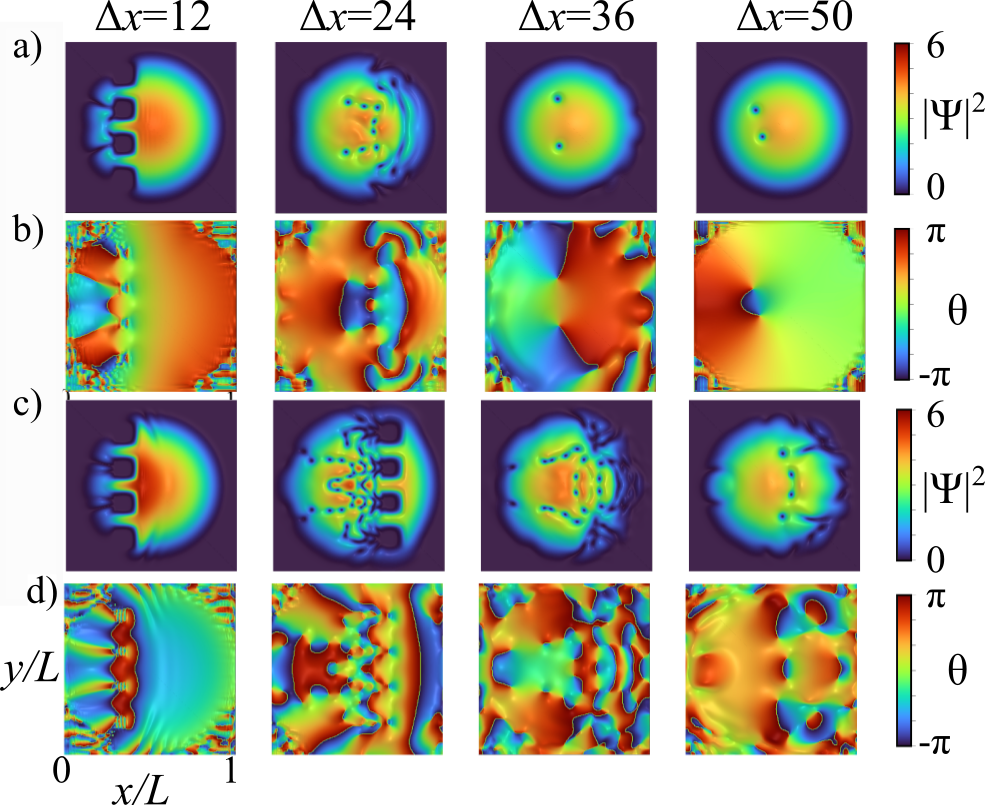}
    \caption{Snapshots of the dynamics as four paddles are swept through a BEC confined in a harmonic trap with sweep velocity $v=0.5$ ( $\mathrm{a}),\mathrm{b})$) and $v=1.5$  ( $\mathrm{c}),\mathrm{d})$). After the paddles have swept through a distance of $\Delta x= 20$ they begin to be linearly ramped down over a fixed time period of $T_r=5$. We show snapshots of the dynamics after the paddles have swept through various distances $\Delta x$, where both  $\mathrm{a}),\mathrm{c})$ show the density $|\Psi|^2$ and $\mathrm{b}),\mathrm{d})$ the phase $\theta$. We use simulation parameters of $L=35,g=1,\gamma=0.1,\Delta t=0.0005$ on a $256\times 256$ spatial grid, with a harmonic trap frequency of $\omega=0.15\sqrt{2}$ }
    \label{fig:Harmonic_v=0.5}
\end{figure}

\begin{figure}[t]
    \centering
    \includegraphics[width=\linewidth]{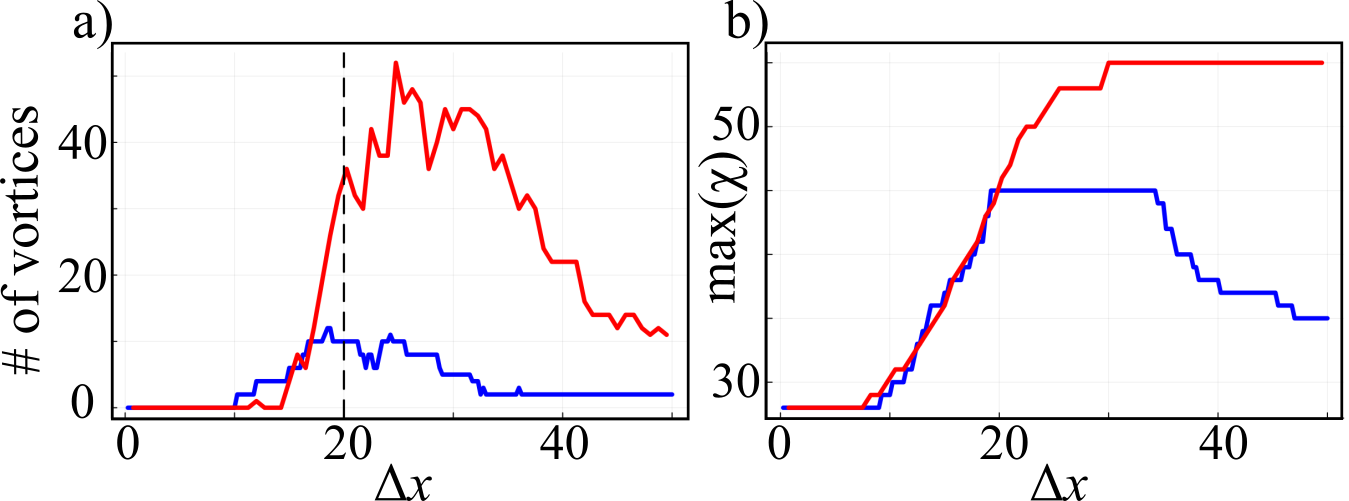}
    \caption{$\mathrm{a})$ Number of vortices present in simulation domain behind the paddles and $\mathrm{b})$ maximal bond dimension of $\Psi$ as a function of the distance the paddles swept through ($\Delta x$) for the simulation shown in Fig.~\ref{fig:Harmonic_v=0.5}. After the paddles sweep through a distance $\Delta x=20$ (black dashed line), they are ramped down to zero over a time of $T_r=5$, and system left to freely evolve. Shown for a paddle sweep velocity of $v=0.5$ (blue) and $v=1.5$ (red).}
    \label{fig:Harmonic_vortices}
\end{figure}

We also investigated an alternative time evolution scheme widely used for MPS in the form of the time-dependent variational principle (TDVP). TDVP is a local time evolution scheme, which sweeps across the MPS, evolving each tensor individually under the action of the local effective Hamiltonian \cite{TDVP}. In Appendix~\ref{app:tdvp}, we further outline the TDVP approach and show that while TDVP can provide smaller errors than those of the finite difference and split-step method for large cutoffs, but larger errors for smaller cutoffs.

The maximal bond dimension to represent the function $\Psi$ was also tracked throughout the simulations, and displayed in Fig.~\ref{fig:dt=0.001_shed_Bdims}. Across all simulation methods, we observe modest bond dimensions which depend on the truncation cutoff chosen. When simulating with a truncation cutoff of $10^{-15}$, the observed maximal bond dimension appears to grow linearly with time as the dynamics unfold reaching a peak value of 35 at the end of the simulation. Using a smaller truncation cutoff results in lower maximal bond dimensions, which increases at a reduced rate, as illustrated in Fig.~\ref{fig:dt=0.001_shed_Bdims}b and c.

We now consider how the speed of the paddles impacts the ability to represent the solution efficiently in MPS format. In general, it is observed that the faster the paddle sweeps through the BEC, the more vortices are generated, and as the velocity is increased further still, we begin to see the emergence of waves resembling shock waves, which can be observed in Fig.~\ref{fig:vel_snaps}. 

A series of paddle velocities were considered, and the corresponding maximal bond dimensions during the simulations are also shown in Fig.~\ref{fig:vel_snaps}. It appears that increasing the velocity of the paddle in general requires a growth in the bond dimension to represent the resultant dynamics. Increasing the sweep velocity generates a lot more shock fronts and produces many more vortices behind, resulting in a much less smooth solution compared to a low sweep velocity, thus requiring a larger bond dimension to encode.

The ability to restrict tensor network simulations to low bond dimensions is ultimately what provides the benefit over other simulation techniques. For a fixed bond dimension, the computational complexity of the tensor network methods scales linearly in the length of the MPS/MPO, corresponding to a logarithmic scaling with number of grid points, and only polynomially in terms of the bond dimension. This yields an overall computational scaling as $\mathcal{O}(\chi^4 \log{M})$, with the dominant contribution from the calculation of the non-linear term $|\Psi|^2$. This contrasts with the corresponding complexity scaling of direct numerical simulations, which scale according to $\mathcal{O}(M \log{M})$ with the dominant scaling via that of the fast fourier transform. To demonstrate this scaling of the tensor network method, we calculated the required CPU time to perform the vortex shedding simulation outlined above across various grid-size, ranging from $64\times 64$ up to $1024 \times 1024$, for different fixed bond dimensions. Results are illustrated in Fig.~\ref{fig:shed_bench}a. We find precisely a linear scaling in the CPU time required with the length of the MPS, ranging from $10^{4}$s on a $64 \times 64$ spatial grid to $10^{5}$s on an $1024\times 1024$ spatial grid, with a bond dimension of $\chi=32$.

Along with the speedup, there is also the data compression aspect. By restricting $\chi$, we reduce the number of terms one requires to represent the state $\Psi$, reducing the memory requirements to simulate the dynamics. One can calculate the ratio of the number of terms to encode $\Psi$ as an MPS with given bond dimension, to the number of terms required to represent $\Psi$ during a direct numerical simulation. This ratio decreases exponentially fast as one increases the length of MPS, as observed in Fig.~\ref{fig:shed_bench}b. Combined, the data compression and scaling of tensor networks allows one to exploit these methods to simulate incredibly fine spatial resolutions, which would be otherwise unfeasible with direct numerical simulations. 

\begin{figure*}[t]
    \centering
    \includegraphics[width=\linewidth]{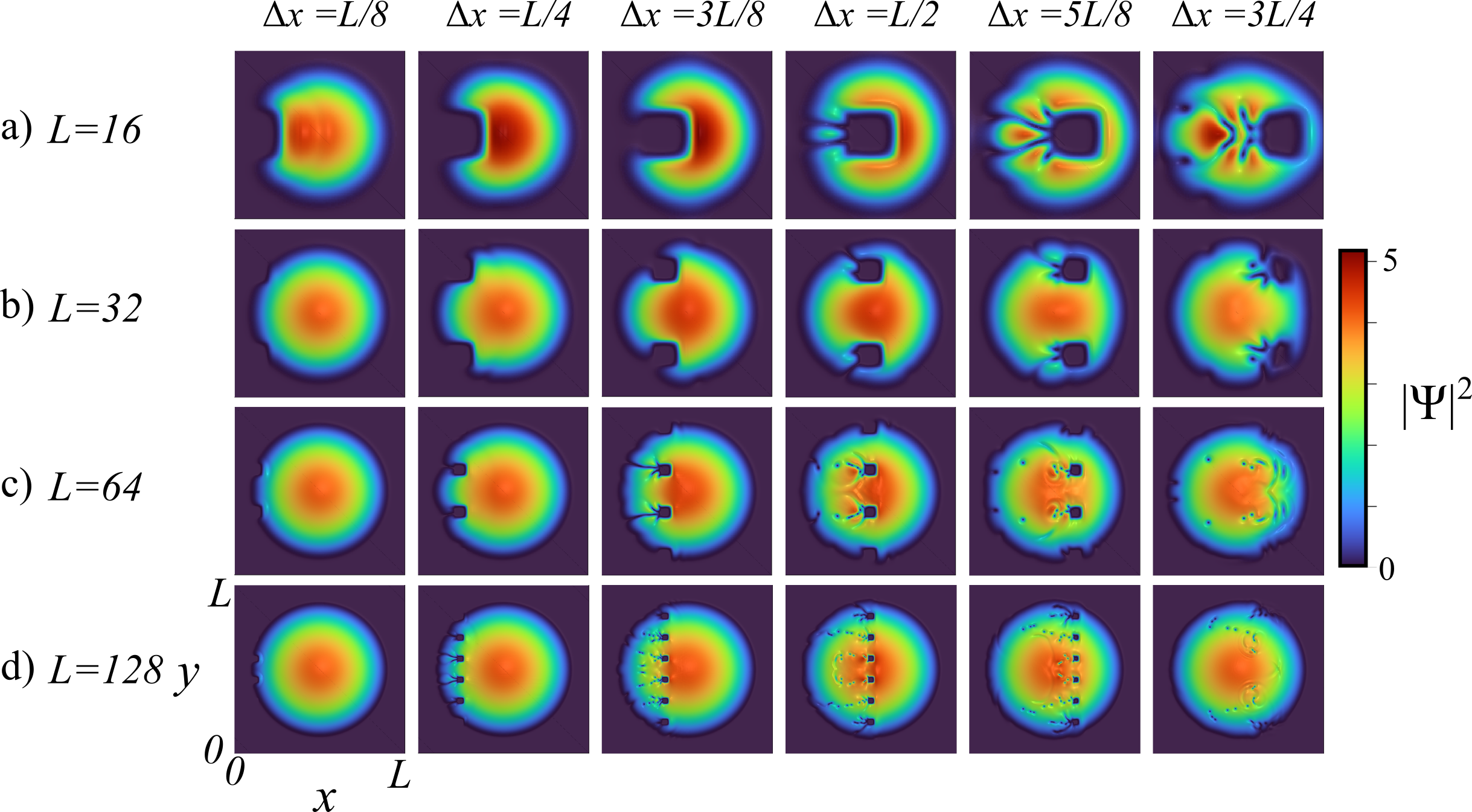}
    \caption{Snapshots of vortex formation after sweeping paddles through increasingly large harmonically trapped BECs, increasing the simulation width from $L=16$ in $\mathrm{a)}$ all the way up to $L=128$ in $\mathrm{d)}$. In each simulation, we maintain a fixed density of grid-points of $L^2/M = 0.015625 $, where $M$ is the total number of grid-points in the simulation, and a fixed paddle width $w=5$. As we increase the system size $L$, we also increase the number of uniformly spaced paddles present in the simulation, all moving with a fixed sweep velocity of $v=1.0$. We plot the condensate density $|\Psi|^2$ after the paddles have swept through various distances $\Delta x$ for the following parameters; $\gamma=0.1,\Delta t=0.005,g=1$, and a truncation cutoff of $10^{-12}$. }
    \label{fig:Harmonic_scale}
\end{figure*}

\subsubsection{Multiple paddles}

Having demonstrated and bench-marked the ability of one to use tensor networks to perform simulations of the GPE in the previous sections, we now use our tensor network approach to further probe vortex dynamics and interactions in the modified GPE. We consider an alternative stirring mechanism, replacing the one large paddle with four smaller paddles, all aligned along the y direction. This type of geometry has been considered previously in experiments \cite{Giant_Vortex_Gauthier_2019}, and allows more vortices to be injected into the BEC for a given sweep velocity at a given time. We introduce the paddles into the BEC, which are then swept across in the $x$ direction with a constant sweep velocity $v$ as was the case previously. Once the paddles have travelled through halfway across the BEC, we linearly ramp down the potential strength to zero over a fixed time $T_{r}$, after which the vortices are left to evolve freely.
In addition, we introduce a fixed harmonic trap throughout the simulation to confine the BEC into a small region, allowing us to trap the created vortices. We consider a fixed dissipation rate of $\gamma=0.1$, and two sweep velocities $v=0.5$ and $v=1.5$,  with snapshots of the resultant dynamics illustrated in Fig.~\ref{fig:Harmonic_v=0.5}.

One can observer vortices continuously shedding off the paddles as they sweep across the BEC, with more vortices generated behind the faster moving paddles. The number of vortices during the simulation can also be recorded \cite{Vortex_detection_PhysRevA}, as shown in Fig.~\ref{fig:Harmonic_vortices}a, where we clearly observers that the $v=1.5$ paddle produces around five times the number of vortices as compared to the $v=0.5$ simulation. For both sweep velocities, there is an initial increase in the number of vortices as the paddles move into the bulk of the harmonic trap. After the paddles have swept through a distance of $\Delta x=20$, they are quickly ramped down, and more vortices are created during this stage.

\begin{figure*}[t]
    \centering
    \includegraphics[width=\linewidth]{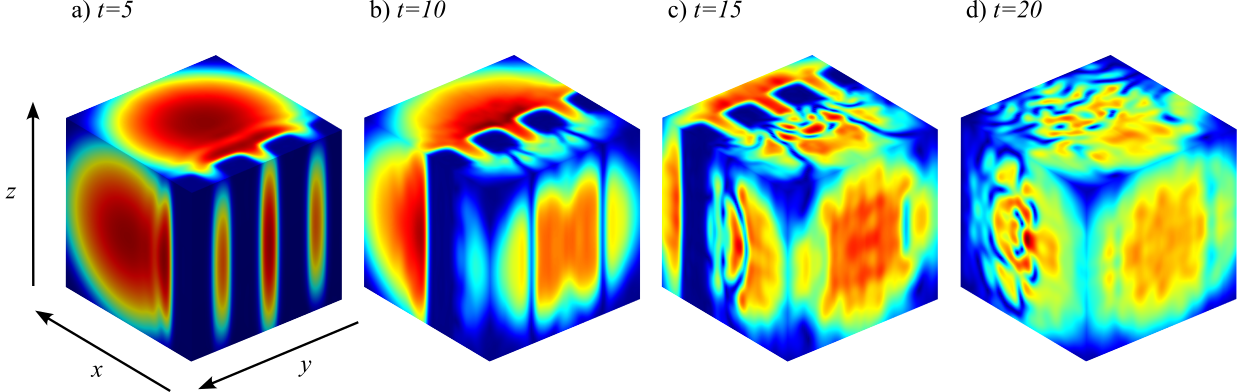}
    \caption{Snapshots of 3D vortex shedding simulation run with a fixed bond dimension of $\chi=32$. We use identical paddles to the 2D simulations, but uniformly extended along the $z$ direction. Snapshots of the dynamics are shown for various times $t$. The simulation is run with the following parameters, $L=35,\omega=0.18 \sqrt{2}$,$\Delta t=0.01$,$\gamma=0.1$,$g=1$ and $v=1.5$ on an $128 \times 128 \times 128$ spatial grid. }
    \label{fig:3D_Bmax=32_3D}
\end{figure*}

\begin{figure*}[t]
    \centering
    \includegraphics[width=\linewidth]{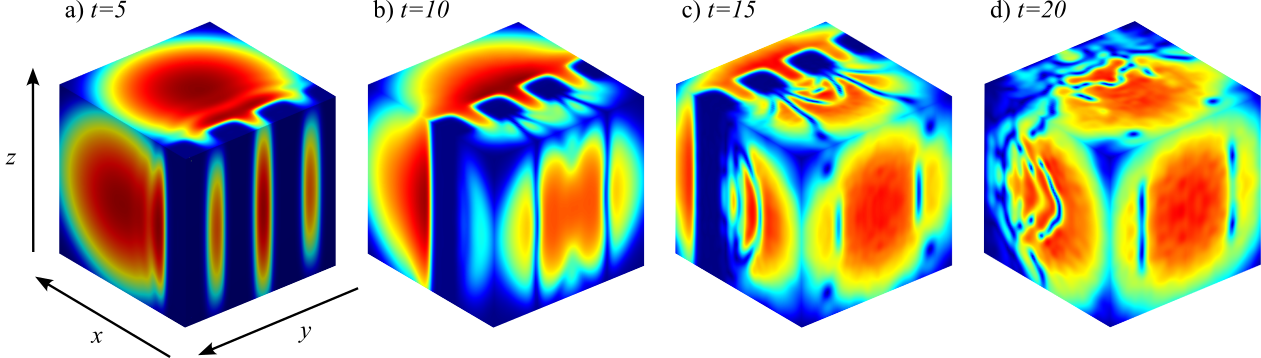}
    \caption{Snapshots of 3D vortex shedding simulation run with a fixed bond dimension of $\chi=64$. We use identical paddles to the 2D simulations, but uniformly extended along the $z$ direction. Snapshots of the dynamics are shown for various times $t$ and at different cuts along the z axis. The simulation is run with the following parameters, $L=35,\omega=0.18 \sqrt{2}$,$\Delta t=0.01$,$\gamma=0.1$,$g=1$ and $v=1.5$ on an $128 \times 128 \times 128$ spatial grid.}
    \label{fig:3D_Bmax=64_3D}
\end{figure*}

After a vortex is formed, it remains present in the simulation until it collides with another vortex with opposite rotational direction, where they annihilate one another. These vortex collisions give rise to a decay in the number of vortices after the paddles are ramped down.
These simulations were conducted using a truncation cut-off of $10^{-12}$, and the resultant maximal bond dimensions are displayed in Fig.~\ref{fig:Harmonic_vortices}b. Again it was found a small bond dimension was required to represent the resultant dynamics, where the bond dimension initially grows linearly in time as the paddles sweep across and generate vortices, before saturating for later times.

To illustrate the true benefit of using tensor networks, we consider performing simulations over increasingly large spatial sizes $L$, whilst maintaining a fixed paddle width ($w$), each time doubling the length of the system $L$ and hence the number of total grid points per dimension, whilst retaining a fixed spatial discretisation $\Delta x$. The following parameters are chosen, $\Delta x=0.125 $, $\Delta t= 0.0005$, $g=1$, $\omega=\frac{5.25\sqrt{2}}{L}$,$\gamma=0.1$, $w=5$ and $v=1.0$, where the harmonic trap frequency is scaled with the system length.

In this way we demonstrate the ability of tensor network simulations to resolve increasingly fine length scales without the exponential increase in memory requirements. 

\subsection{3D Simulation}

As a final example of vortex shedding, we extend the 2D simulations to a full 3D simulation. Once again, we prepare a BEC trapped in a harmonic trap $V(x,y,z)=\frac{1}{2}\omega^2(x^2+y^2+z^2)$, where paddles are swept through with a fixed sweep velocity. For simplicity, we consider the case of translatioanlly invariant paddles along the $z$ axis, i.e. $V_{\mathrm{3D pad}}(x,y,z,t)= V_{\mathrm{2D pad}}(x,y,t)$, where $V_{\mathrm{2D pad}}(x,y,t)$ is the 2D paddle potential as used in the previous section.

As a demonstration, we perform the 3D simulations on an $128 \times 128 \times 128$ spatial grid with fixed bond dimensions of $\chi= 16,32$ and $64$, with snapshots of the dynamics illustrated in Fig.~\ref{fig:3D_Bmax=32_3D} and \ref{fig:3D_Bmax=64_3D}. We see that even for these modest bond dimensions, we are able to clearly resolve the formation of vortices in the 3D BEC. The true scope of the data compression achievable with MPS is most clearly illustrated in 3D. Using a bond dimension of $\chi=64$, one requires only to store $4 \%$ of the number of variables parameterizing the solution compared to DNS.  

These simulations on an $128^3$ spatial grids are beginning to push that which can be achieved easily with DNS \cite{Ashton_turbulence,SMITH2022108314}. Performing a DNS on an $256^3$ spatial grid would require a factor of $8$ growth in the memory requirements, and a corresponding increase in computation time. In contrast, performing a simulation on a $256^3$ spatial grid and beyond, with MPS and a fixed bond dimension, requires only a small increase in computational complexity, which can allow for very dense simulations.

\section{Simulating Dipolar gases with tensor networks}

\begin{figure}[t]
    \centering
    \includegraphics[width=.6\linewidth]{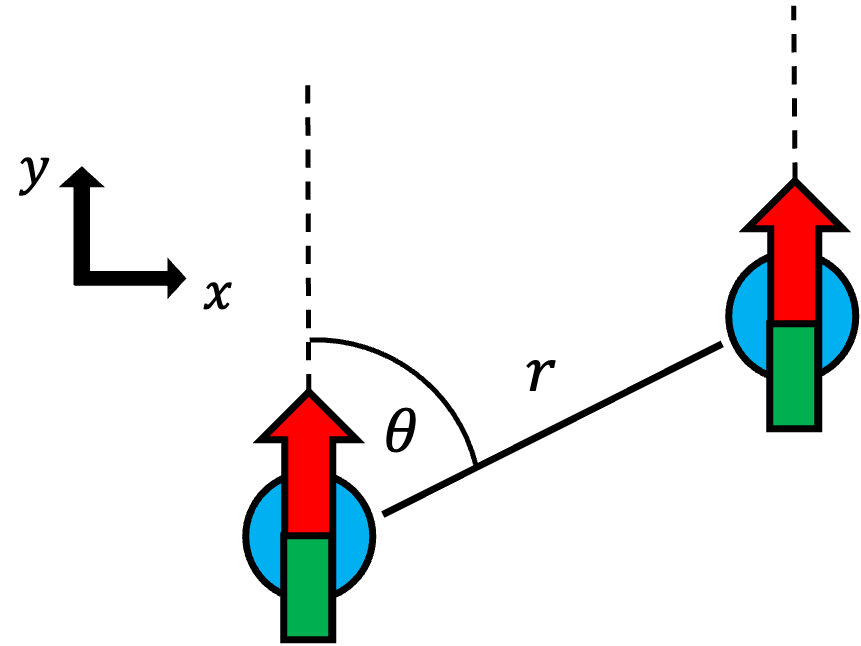}
    \caption{Schematic of two interacting dipoles within the BEC, both with the dipole moments aligned along the polarisation axis $y$. The interaction between these two dipoles is dependent on the separation between them ($r$), and the relative angle between them ($\theta$), as measured from the polarisation axis.}
    \label{fig:dipole_interaction}
\end{figure}

\begin{figure}[t]
    \centering
    \includegraphics[width=\linewidth]{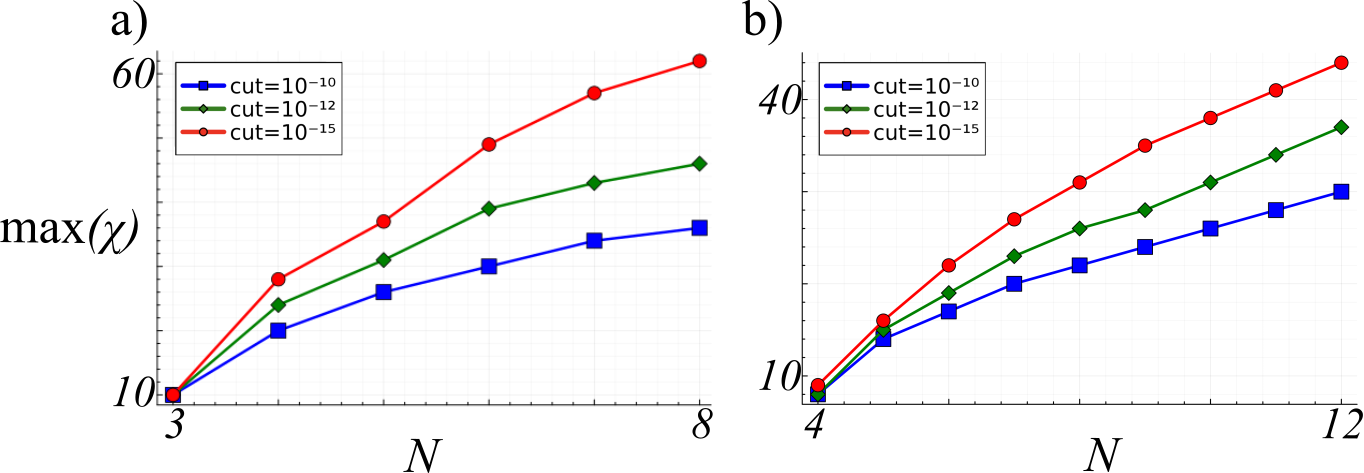}
    \caption{Maximal bond dimension scaling of $U_{dd}$ (Eq.~\ref{eq:Udd_k}) for increasing MPS size ($N$) for truncation cutoffs of $10^{-10}$ (blue squares), $10^{-12}$ (green diamonds) and $10^{-15}$ (red dots). Results are shown for the a) 3D interaction kernel and b) the 2D interaction kernel assuming a uniform $z$ direction. Further details are provided in Appendix.~\ref{app:dipole}. }
    \label{fig:dipole_dims}
\end{figure}

\begin{figure*}[t]
    \centering
    \includegraphics[width=\linewidth]{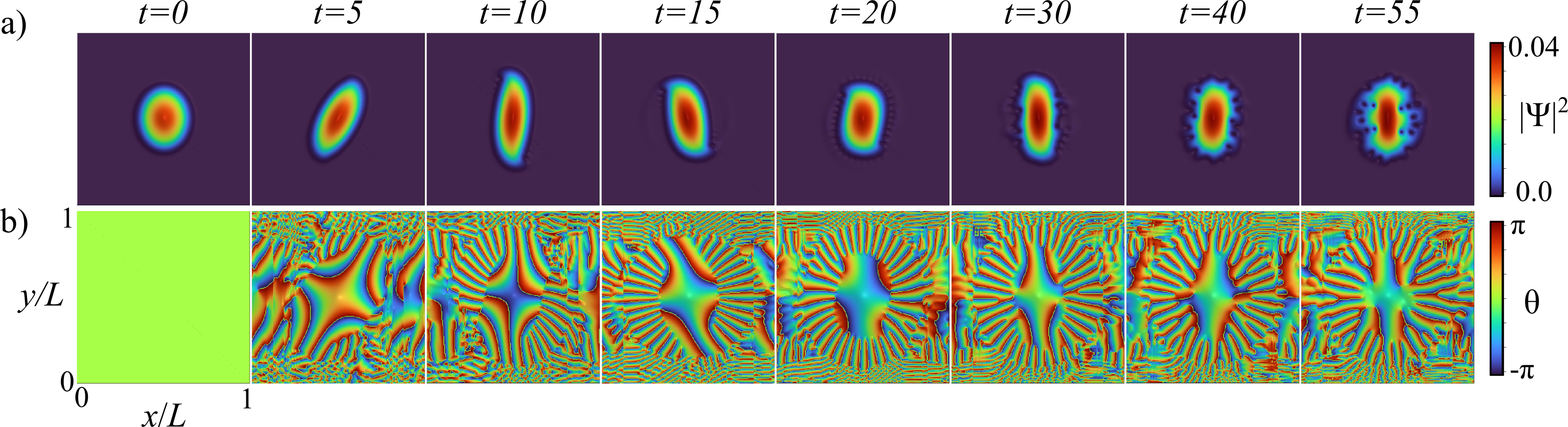}
    \caption{Snapshots of the resultant dynamics of a dipolar gas with parameters $g=250$, $g_{dd}=75, \gamma=0.03$ and $L=25$ as it begins to rotate around the $z$ axis. The condensate is first prepared in the ground state of a harmonic trap with $\omega=1$ ($t=0$), before a sudden rotation at frequency $\Omega=0.7$ is imparted. The simulation is conducted on an $256\times256$ spatial grid, using a time-step of $dt=0.001$ and truncation cutoff of $10^{-15}$. We display both $|\Psi|^2$ and the phase at various simulation times $t$.}
    \label{fig:dipolar_gdd=100_snaps}
\end{figure*}

\begin{figure}[t]
    \centering
    \includegraphics[width=\linewidth]{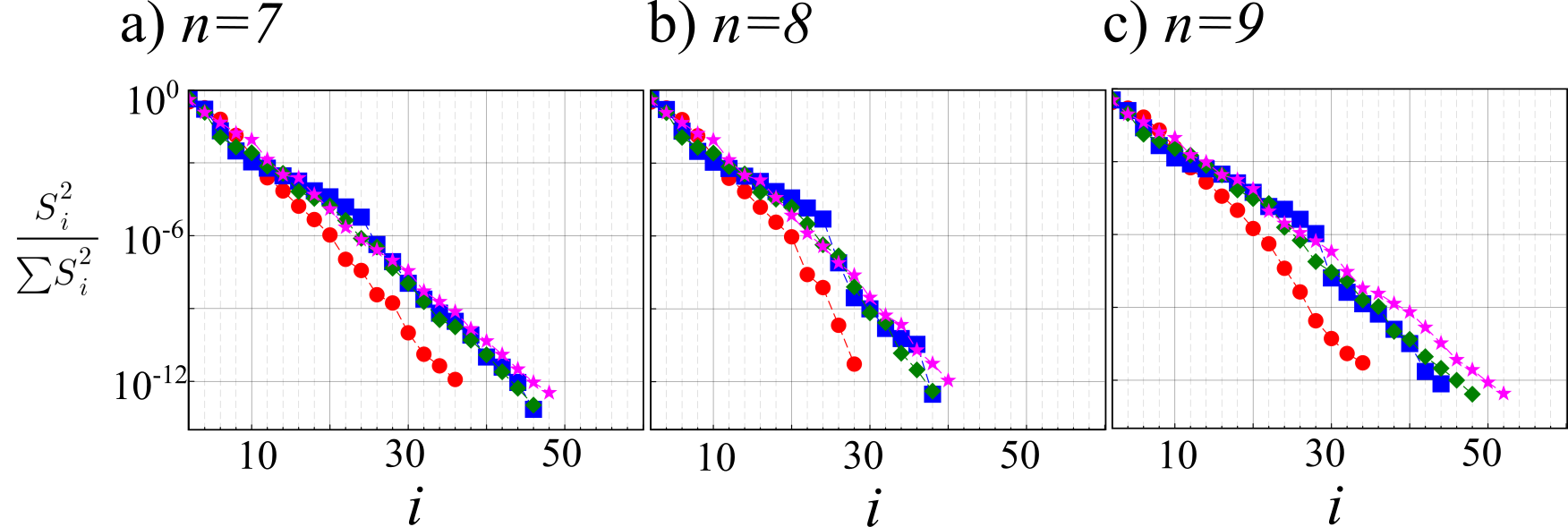}
    \caption{Squared singular values of the MPS $\Psi$ in descending order during the simulation of dipole rotation as outlines in Fig.~\ref{fig:dipolar_gdd=100_snaps}. We plot the singular values across the a)$n=7$ b)$n=8$ and c) $n=9$ bond of the MPS at simulation times of $t=10$ (red dots), $t=25$ (blue square), $t=40$ (green diamond) and $t=55$ (pink stars). At all times and at all bonds the singular values decay linearly in the above plots, indicating the singular values decay exponentially fast. }
    \label{fig:dipolar_sing_vals}
\end{figure}

We have so far shown how one can efficiently use MPS methods to simulate the standard GPE. We will now introduce additional complexity into the GPE, in the form of long-range interactions of a dipolar gas. A dipolar gas is a BEC made from cold atoms with a dipole moment, where an additional interaction is introduced dependent on the relative separation and orientation of neighbouring dipole moments, as illustrated in Fig.~\ref{fig:dipole_interaction}. Speciffically, we consider the case of a polarised dipolar gas, where the dipole moments of all the atoms point along the same direction, which we take throughout as along the $y$ axis. When the dipoles are aligned (i.e. the two atoms separated along the polarisation axis), there will be an attractive interaction, but if separated against the polarisation angle there will be a repulsive interaction. There exists a magic angle $\theta_m$ at which there is zero dipole-interaction between the atoms. 

We can express the dipole-dipole interaction $\Phi(x)$ at some point $x$ as an integral over all space of the condensate density $|\Psi|^2$ and the dipole interaction potential $U_\mathrm{dd}(r)$
\begin{equation}
    \Phi(x) = \int dx' \  U_\mathrm{dd}(x-x')|\Psi(x')|^2   \ , \label{eq:dipole_term}
\end{equation}
with the full 3D potential given by
\begin{equation}
   U_\mathrm{dd}(r) = \frac{3}{4 \pi r^3} (1-\frac{3y^2}{r^2})  \ . \label{eq:dipole_U}
\end{equation}
This dipole-dipole interaction term is introduced into the GPE as follows
\begin{equation}
    i\hbar \frac{\partial \Psi}{\partial t} = \left[ -\frac{\hbar^2}{2m} \nabla ^2 + V(x) + g|\Psi|^2 + g_\mathrm{dd}\Phi(x) \right]\Psi \  ,
\end{equation}
where $g_{dd}$ is known as the dipolar interaction strength and controls the dipolar interactions within the system. One can then apply all the aforementioned MPS based schemes to simulate the dipolar GPE, provided that the dipole-dipole term $\Phi(x)$ can be efficiently constructed in MPS form.

\subsection{Constructing the interaction term}
One possibility to construct $\Phi(x)$ is to numerically intergrate Eq.~\eqref{eq:dipole_term} directly. However, the dipole interaction potential $U_\mathrm{dd}(r)$ is highly singular as it diverges as $r\rightarrow 0$, creating difficulties for numerical integration techniques. There are some approaches to deal with this in real space \cite{TANG2017223}, however, one may reformulate the interaction as a convolution \cite{Smith_2023}
\begin{equation}
   \Phi(\mathbf{r}) = \boldsymbol{\mathcal{F}}^{-1} \{\tilde{U}_\mathrm{dd}(\mathbf{k}) \boldsymbol{\mathcal{F}}\{|\Psi(\mathbf{r})|^2\} \} \ ,
\end{equation}
where $\tilde{U}_\mathrm{dd}(k)$ is the k-space interaction kernel, found via the Fourier transform of the dipole interaction potential (Eq.~\eqref{eq:dipole_U}), and is given by
\begin{equation}
    \tilde{U}_\mathrm{dd}(k)=3\frac{k_y^2}{k^2}-1 \ . \label{eq:Udd_k}
\end{equation}

One can also obtain expressions for $\tilde{U}_\mathrm{dd}(k)$ for effective 2D dipolar systems, where one makes assumptions about the behaviour along the $z$ axis. One could assume the condensate is uniform along the z direction, in which case $U_\mathrm{dd}(k)$ takes on the same form as Eq.~\eqref{eq:Udd_k}. Alternatively one could also consider a condensate in a strong harmonic trap along the $z$ direction, where all dynamics occur in the x-y plane only. Details of the 2D dipolar interactions are provided in Appendix.~\ref{app:dipole}

We note that calculation of the dipolar interactions via the Fourier transform implies the presence of periodic boundaries. This will have an unwanted effect of including in the dipolar terms the contribution from copies of the physical system outwith the simulation domain. To mitigate this, one will often apply a cutoff in the dipolar interaction kernel to avoid these unphysical contributions. We outline the details of this and its tensor network implementation in Appendix \ref{App:momentum_cutoff}.

To simulate dipolar gases with the MPS approach, $\tilde{U}_\mathrm{dd}(\mathbf{k})$ is first created as an array for all $\mathbf{k}$ points, and SVDs are then performed to encode into the correct MPS form, where we then truncate with some truncation cutoff.  The resultant maximal bond dimensions are illustrated within Fig.~\ref{fig:dipole_dims}. The maximal bond dimension does grow with increasing system size as the bond dimension appears to scale logarithmically with MPS length, and not exponentially. This would thus allow one still to push simulations to very fine spatial resolutions without a blow-up of bond dimension to represent the interaction kernel.

\subsection{Vortex formation upon rotation}

A further test case for MPS-based simulation methods is now considered where we study the formation of vortices as we begin to spin the condensate within a harmonic trap \cite{Carlos_lobo_vortex,PhysRevA.65.023603,PhysRevA.67.033610,10.1063/5.0143556,vortex_stripes_dipolar,PhysRevA.100.023625}. A dipolar BEC is trapped in a stationary, non-rotating harmonic trap. Then a sudden rotation at frequency $\Omega$ is imparted on the trap and a small anisotropy is introduced to impart angular momentum onto the BEC. To account for this rotation, we introduce the angular momentum operator $L_z=-i\hbar(x\partial_y -y\partial_x)$ into the dipolar GPE as follows
\begin{equation}
\begin{aligned}
    i\hbar \frac{\partial \Psi}{\partial t} = \Bigg[ & -\frac{\hbar^2}{2m} \nabla ^2 + V(x) -\mu + g|\Psi|^2 + g_\mathrm{dd}\Phi(x)  \\ & -\Omega L_z \Bigg]\Psi \  . 
\end{aligned}
\end{equation}
For the simulations which follow, we consider a strong conventional interaction term of $g=250$ and dipolar interaction strength of $g_\mathrm{dd}=75$, where we first use imaginary time-evolution to prepare the groundstate of the stationary trap, before the rotation is suddenly turned on to $\Omega=0.7$ . Snapshots of the resultant dynamics are displayed in Fig.~\ref{fig:dipolar_gdd=100_snaps}, displaying both the density of the BEC and the associated phase. 

The condensate starts in a rotationally symmetric state at $t=0$, but as we begin to spin the trap, we observe the condensate morph into a more elliptical shape which rotates around. After an initial period, we begin to observe the formation of vortices at $t\sim 40$, which can again be identified via the quantised circulation of phase. The associated singular values for these dynamics are displayed in Fig.~\ref{fig:dipolar_sing_vals} across different bonds $n$ of the MPS, and for various simulation times. There is a clear exponential decay of singular values across all times, which implies the ability of one to truncate the MPS bond dimension substantially, by discarding the least significant singular values.%

In this paper we have demonstrated the ability to perform simulations of the GPE on fine grids making use of the tensor network formalism, where the main advantage is that of data compression. Note that this is expected to work well for cases where those fine grids are necessary, but reformulating a direct simulation in terms of tensor networks introduces some computational overhead. Therefore if the grid sizes required for some problem can be handled with a more conventional direct solver, then conventional solvers would often be a better approach as opposed to using a tensor network formalism. We have noted several examples in the manuscript involving turbulence and dipolar interactions where fine grids may be required, and therefore tensor networks can give an advantage. Furthermore, even for those problems were tensor networks may be of benefit, the data compression one can achieve with tensor networks for a given problem will depend upon the physics of that instance, in the form of the structure of the correlations present. Therefore one must test that tensor networks can provide useful data compression for each instance of a problem.

In addition, we note that other methods have been developed which can further compete against tensor network approaches such as neural networks (NN). NNs have also been applied to the study of the GPE, where they have been demonstrated to be able to solve a $1D$ time independent GPE for a variety of external potentials.  The drawback with NNs as compared with tensor-network approaches is that they require sufficient training of the network to reproduce accurate results, which may be computationally intensive. But in general NNs have been used in the past for other partial differential equations \cite{712178}, such as for simulations of fluid dynamics \cite{cai2021physicsinformedneuralnetworkspinns} and heat transfer problems, where a physics-informed neural network (PINN) was used to qualitatively capture the dynamics of a convection flow past a cylinder, with errors on the order of a few percent level as compared with a direct solver \cite{10.1115/1.4050542}. Such NN methods may also be able to be applied for dynamical simulations of the GPE and dipolar gases. 

\section{Conclusion}

We have shown how MPS can be exploited to encode solutions of the GPE and achieve substantial data compression for dynamics where we are able to restrict the bond-dimension. Along with finite-difference MPOs to calculate derivative terms, the quantum Fourier transform provides a useful alternative way to perform time-evolution dynamics, often producing smaller errors for the same bond-dimension as compared with the finite-difference counterpart. Going beyond the GPE, we included effects of dipolar interactions, again relying on the quantum Fourier transform. We demonstrated how the required bond dimension of the interaction term scales linearly in MPS length, and thus one may still be able to achieve an exponential data compression as compared with direct simulations. This work opens up the possibility of performing extremely finely resolved numerical simulations of cold gases, beyond that which would be otherwise possible. Having access to very fine spatial resolution is key for studying and understanding quantum turbulence.

We have so far focused on time-evolution of single component bose gases. However, future work may begin to explore the dynamics of multi-component systems, such as a binary condensate or a spin-1 system. By modifying the interactions between components, one can explore a variety of rich physics and explore phase transitions \cite{SMITH2022108314,Hern_ndez_Rajkov_2024,PhysRevLett.116.025301} . In addition, we have here demonstrated inclusion of dipolar interactions. An important future avenue of work may be a further detailed  exploration of dipolar physics, such as the formation of super-solids \cite{poli2024excitationstwodimensionalsupersolid,PhysRevA.108.053321,Bland_2024,PhysRevA.109.023313}, while exploiting MPS to perform simulations on very fine spatial grids.

\section*{acknowledgements}
 We thank Paula Garc\'ia-Molina, Preetma Soin, Neil Gaspar, Ashton Bradley, Blair Blakie and Dieter Jaksch for helpful discussions. This work was supported by the EPSRC through the Programme Grant QQQS (EP/Y01510X/1), and grant EP/Y005058/2. The authors would like to acknowledge AWE plc for financial support of the work.
UK Ministry of Defence © Crown owned copyright 2024/AWEs
\bibliographystyle{apsrev4-2}
\bibliography{Draft}

\appendix
\section{Time evolution schemes}

\subsection{4th order Runge-Kutta} \label{App:RK4}
A frequently used numerical scheme to perform time evolution of differential equations is that of Runge-Kutta 4 (RK4). RK4 belongs to a class of Runge-Kutta solvers, however RK4 is most widely used for its good accuracy and stability for solving PDEs.

Given a PDE of the following form,
\begin{equation}
    \frac{\partial \Psi(t)}{\partial t} = -\frac{i}{\hbar} \hat{H}(t) \Psi(t) \ ,
\end{equation}
one performs RK4 via evaluating the $ \frac{\partial \Psi(t)}{\partial t}$ at 4 intermediate points in order to produce $\Psi(t+dt)$. Specifically, one must calculate the following intermediary states during an RK4 time-step
\begin{eqnarray}
    K_1 &=& \frac{\partial \Psi(t)}{\partial t}  , \\ 
    \Psi_1 &=& \Psi(t) +\frac{dt}{2} K_1 , \\
    K_2 &=& \frac{\partial \Psi_1(t+\frac{dt}{2})}{\partial t} , \\
    \Psi_2 &=& \Psi(t) +\frac{dt}{2} K_2 , \\
    K_3 &=& \frac{\partial \Psi_2(t+\frac{dt}{2})}{\partial t} , \\
    \Psi_3 &=& \Psi(t) + dt K_3 ,\\
     K_4 &=& \frac{\partial \Psi_3(t+dt)}{\partial t} .
\end{eqnarray}
Having obtained the intermediate derivatives $K_1,K_2,K_3$ and $K_4$, we produce the next time-step in the simulation via addition with the solution $\Psi(t)$ at the current time,
\begin{equation}
    \Psi(t+dt) = \Psi(t) +\frac{dt}{6}(K_1+2K_2+2K_3+K_4) \ .
\end{equation}

Specifically for our implementation with tensor networks, the derivatives $K_i$ are obtained via an MPO-MPS contraction with the Hamiltonian MPO $\hat{H}(t)$ and the MPS representing $\Psi$. One can then add a number of MPS in the usual manner \cite{DMRG} to produce the MPS for the evolved solution.

\subsection{TDVP evolution} \label{app:tdvp}

A further evolution scheme available to us is that of time dependent variational principle (TDVP). TDVP is a local time-evolution method which evolves an MPS under the action of some Hamiltonian $H$. TDVP works in a similar fashion to the famous DMRG algorithm, where one evolves the first MPS tensor in time by a small amount $dt$ under a local effective Hamiltonian, before sweeping right across the MPS evolving each tensor and bond in a similar fashion \cite{TDVP,Paeckel_2019}. We implement TDVP by using the code package ITensorTDVP \cite{ITensors}, which works with the ITensors library to allow for efficient time-evolution.

To evolve the GPE using TDVP, we first must construct the Hamiltonian as an MPO. The Hamiltonain can be separated as ,
\begin{equation}
    H= -\frac{1}{2}\nabla^2 + H_d \ , 
\end{equation}
where $H_d = V(\mathbf{r}) -\mu + g|\Psi|^2$ is a diagonal MPO in real space. One can construct the Laplacian term using both finite difference methods or by exploiting the QFT as follows,
\begin{equation}
    \nabla^2 =  - \  \boldsymbol{\mathcal{F}^{-1}} \sum_{\mathbf{k}} \left(k_x^2 + k_y^2 + k_z^2\right) \ket{\mathbf{k}}\bra{\mathbf{k}} \boldsymbol{\mathcal{F}} \ .
\end{equation}

We found that TDVP performed using the QFT Laplacian produces smaller errors as compared to the finite difference Laplacian across all times in the simulation, and for all truncation cutoffs. For this reason, whenever using TDVP for full time evolution of the GPE, we  constructed the Hamiltonian using the QFT based Laplacian. 

\begin{figure}[htb!]
    \centering
    \includegraphics[width=\linewidth]{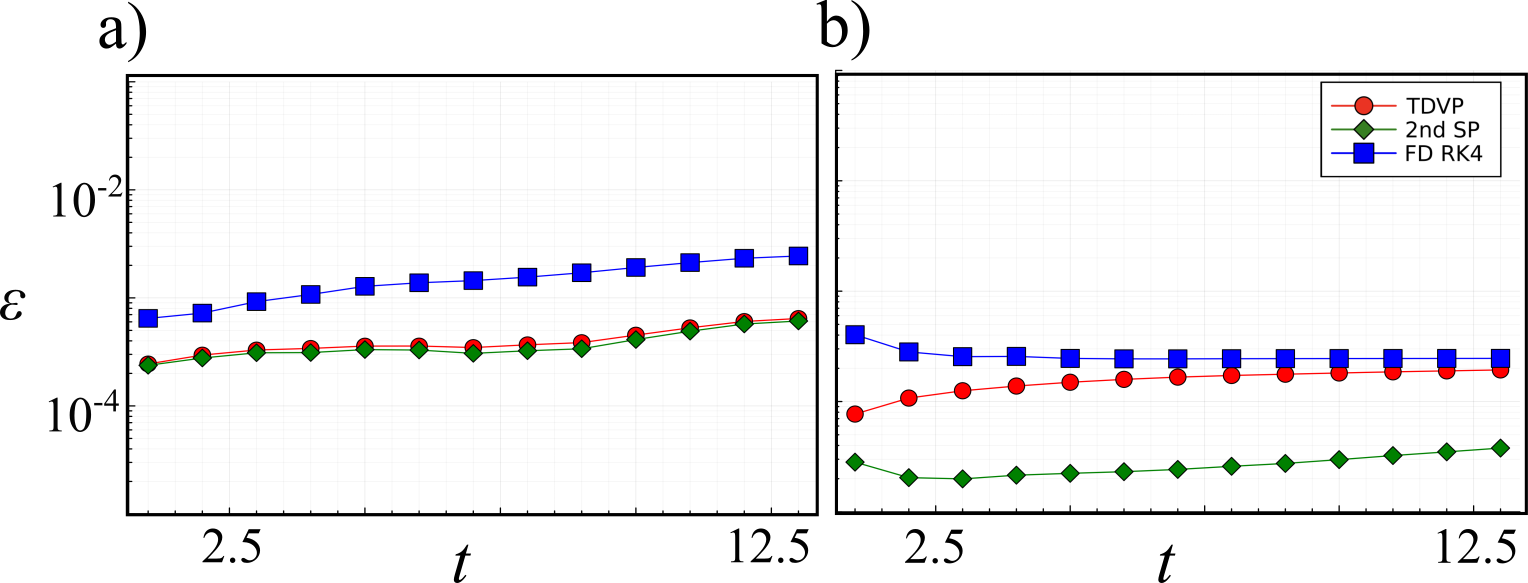}
    \caption{$\mathrm{a})$ Errors from MPS based vortex shedding simulations against XMDS2 direct simulation, for an $256\times 256$ spatial grid, with $\gamma=0.1, v=1.5$ and $\Delta t=0.001$. We compare errors during time evolution using the finite difference RK4 evolution scheme (blue squares), a second order split-step scheme (red dots) and using TDVP  (green diamonds). The MPS simulations are run for different truncation cutoffs of $\mathrm{a})$ $10^{-12}$,and $\mathrm{b})$ $10^{-15}$. }
    \label{fig:errs_tdvp}
\end{figure}

We perform an identical simulation as in Sec. \ref{sec:single_paddle} but using TDVP for the full time evolution, with the resultant errors relative to XMDS2 illustrated in Fig.~\ref{fig:errs_tdvp}. We consistently find that we obtain very similar errors for truncation cutoffs of $10^{-12}$ between TDVP and the split-step schemes, however for a truncation cutoff of $10^{-15}$ performing a split-step evolution scheme produces a smaller error. Additionally, we always find that TDVP produces smaller errors compared to the finite difference RK4 scheme.
\section{Paddle generation} \label{App:Rounded_paddles}

\begin{figure}[htb!]
    \centering
    \includegraphics[width=.9\linewidth]{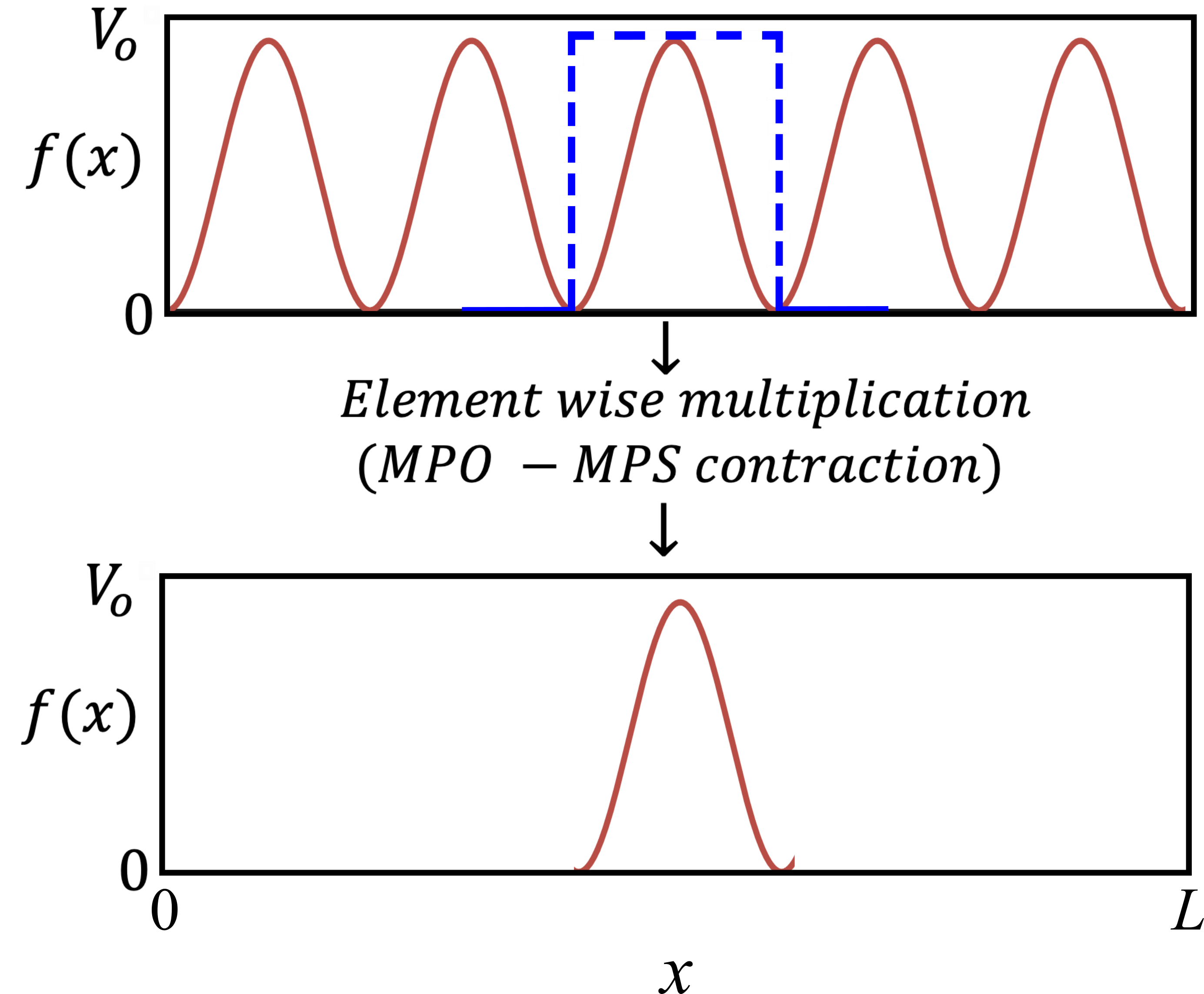}
    \caption{Method for construction of rounded paddles directly in MPS form. An extended function across the whole simulation domain, such as $\sin(x)^2$, is created directly as an MPS, shown in red in the top diagram. A top-hat function is then created as an MPO  (shown as dashed blue line) centered on the desired region of the extended function. One then contracts the MPO with the MPS to produce a localised rounded paddle shape as shown in bottom in diagram.  }
    \label{fig:Rounded_paddles}
\end{figure}

To generate the MPS representation of such paddles, we first construct an MPS representation of the top-hat function, and an MPS representation of an extended rounded potential, e.g. a sine wave. One can then multiply these functions together to create a rounded paddle shape directly in MPS form as depicted in Fig.~\ref{fig:Rounded_paddles}. Note that while this allows for the simulation of the type of strong potentials utilised in recent experiments, we are not implementing any boundary conditions for the quantum fluid at the boundary of the strong potential, e.g., we do not strictly enforce the derivative to go to zero.

\section{2D Dipolar interaction} \label{app:dipole}
Given the full 3D dipolar interaction term, it is possible to construct a reduced 2 dimensional model via assumptions on the condensate along the $z$ directions.  The 3D dipolar interaction term is constructed via a convolution, exploiting the Fourier transform, 
\begin{equation}
   \Phi(\mathbf{r}) = \boldsymbol{\mathcal{F}}^{-1} \{\tilde{U}_\mathrm{dd}(\mathbf{k}) \boldsymbol{\mathcal{F}}\{|\Psi(\mathbf{r})|^2\} \} \ , \label{eq:app_phi}
\end{equation}
where $ \tilde{U}_\mathrm{dd}(k_x,ky_,k_z)$ is the dipole interaction term in momentum space,
\begin{equation}
    \tilde{U}_\mathrm{dd}(k_x,k_y,k_z)=3\frac{k_y^2}{k^2}-1 \ .
\end{equation}

For the following sections, it is important to realise one can split the Fourier transform separately between different dimensions,
\begin{equation}
\begin{aligned}
    \boldsymbol{\mathcal{F}} & = & \frac{1}{(2\pi)^3}\iiint dx \: dy \: dz \: e^{i(k_x x +k_y y +k_z z)} \\ & = & \frac{1}{2\pi}\int \: dz e^{ik_z z} \boldsymbol{\mathcal{F}_{2D}}
\end{aligned}
\end{equation}

\subsection{Translational invariance}

If one assumes the condensate is uniform along the $z$ direction, we can remove the spatial dependence along this direction as,
\begin{equation}
    \Psi(x,y,z)=\psi(x,y) \ .
\end{equation}
Substituting into Eq.~\eqref{eq:app_phi}, and separating the $z$ components of the Fourier transforms, one obtains,
\begin{widetext}
\begin{equation}
   \Phi(\mathbf{r}) = \frac{1}{2\pi} \int dk_z e^{-ikz} \left[ \int e^{ik_z z'} dz' \boldsymbol{\mathcal{F}_{2D}^{-1}} \{\tilde{U}_\mathrm{dd}(\mathbf{k}) \boldsymbol{\mathcal{F}_{2D}}\{|\psi(x',y')|^2\} \} \right] 
\end{equation}
\end{widetext}
The above expression can be reduced by means of the dirac delta function, defined as,
\begin{equation}
    \frac{1}{2\pi}\int dx e^{i(k-k_0)x} = \delta(k-k_0) \ .
\end{equation}
Substituting into the previous expression yields,
\begin{equation}
   \Phi(\mathbf{r}) = \int dk_z \delta(k_z) e^{-ik_z z} \left[\boldsymbol{\mathcal{F}_{2D}^{-1}} \{\tilde{U}_\mathrm{dd}(\mathbf{k}) \boldsymbol{\mathcal{F}_{2D}}\{|\psi(x',y')|^2\} \} \right] , 
\end{equation}
which can be easily simplified to the reduced 2D dipolar model for a translationally invariant condensate along the $z$ axis,
\begin{equation}
   \Phi(\mathbf{r}) = \boldsymbol{\mathcal{F}_{2D}^{-1}} \{\tilde{U}_\mathrm{dd}^{2D}(k_x,k_y) \boldsymbol{\mathcal{F}_{2D}}\{|\psi(x',y')|^2\} \} .
\end{equation}
The 2D dipolar interaction kernel $\tilde{U}_\mathrm{dd}^{2D}$ is thus defined as the usual 3D interaction kernel but evaluated only in the $k_z=0$ plane,
\begin{equation}
    \tilde{U}_\mathrm{dd}^{2D}(k_x,k_y)= \tilde{U}_\mathrm{dd}(k_x,k_y,k_z=0) \ .
\end{equation}

\subsection{Harmonically trapped}
If instead a sufficiently strong harmonic trap is applied along the $z$ axis, then one can assume that the condensate remains in the groundstate of the harmonic trap along the $z$ axis, and the dynamics unfold along the $x-y$ plane \cite{Dipole_harmonic_PhysRevA.82.043623}. One can express  $\Psi$ as 
\begin{equation}
    \Psi(x,y,z)= \psi(x,y)\phi(z) \ ,
\end{equation}
where $\phi(z) = (\frac{m \omega_z}{\hbar \pi})^{\frac{1}{4}} e^{\frac{-m \omega_z}{2\hbar}z^2}$ is the groundstate along the $z$ direction.

Substituting into Eq.~\eqref{eq:app_phi}, one obtains,
\begin{widetext}
\begin{equation}
   \Phi(\mathbf{r}) = \frac{1}{2\pi} \int dk_z e^{-ikz} \left[ \int e^{ik_z z'} \phi(z) dz' \boldsymbol{\mathcal{F}_{2D}^{-1}} \{\tilde{U}_\mathrm{dd}(\mathbf{k}) \boldsymbol{\mathcal{F}_{2D}}\{|\psi(x',y')|^2\} \} \right]  \ .
\end{equation}
\end{widetext}
Following on from this, it is possible to derive the reduced 2D interaction kernel for this problem \cite{Dipole_harmonic_PhysRevLett.106,Dipole_harmonic_PhysRevLett.111}, given by,

\begin{equation}
    U_\mathrm{dd}^{2D}(k_x,k_y) =  -1 +3\sqrt{\pi} \frac{q_y^2}{q^2}e^{-q^2} \mathrm{erfc}(q)
\end{equation}
where $\mathbf{q}=\mathbf{k}\frac{l_z}{\sqrt{2}}$, $l_z=\sqrt{\frac{\hbar}{m \omega_z}}$, and $\mathrm{erfc}(q)$ is the complimentary error function. Note that it is possible to extend this result to the case when the dipolar polarisation axis is aligned along the $z-y$ plane, at some angle $\alpha$ to the $z$ axis \cite{Dipole_harmonic_PhysRevLett.106,Dipole_harmonic_PhysRevLett.111},
\begin{equation}
    U_\mathrm{dd}^{2D}= \cos(\alpha)^2 F_{\perp} +\sin(\alpha)^2 F_{\parallel} \ ,
\end{equation}
where $F_{\perp} = 2-3\sqrt{\pi}qe^{q^2}\mathrm{erfc}(q)$  and $F_{\parallel}= -1 +3\sqrt{\pi} \frac{q_y^2}{q^2}e^{-q^2}$.

\begin{figure}[htb!]
    \centering
    \includegraphics[width=\linewidth]{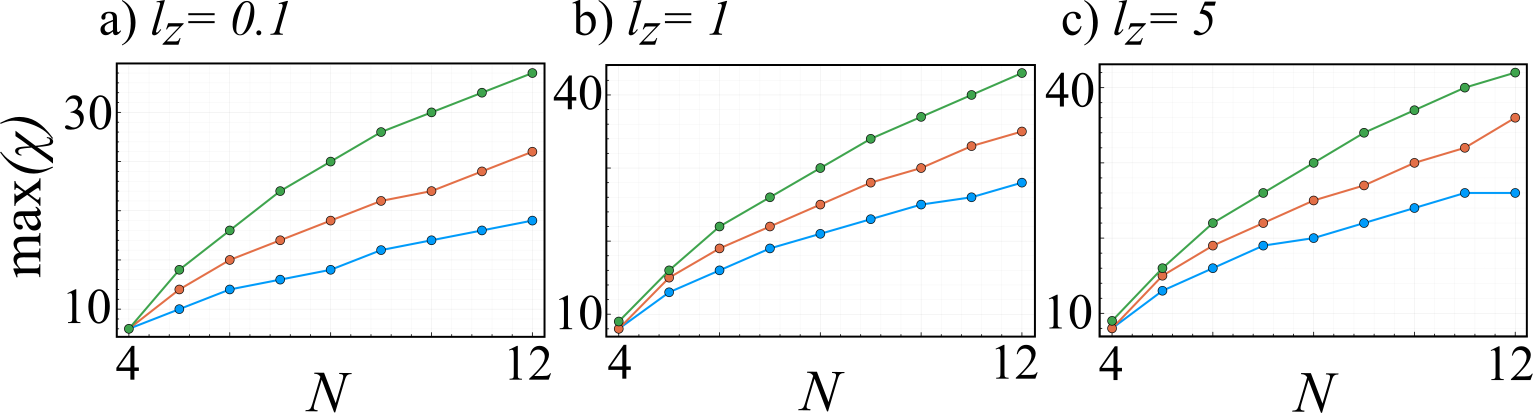}
    \caption{Maximal bond dimension scaling of $U_{dd}$ (Eq.~\ref{eq:Udd_k}) for increasing MPS size ($N$) for truncation cutoffs of $10^{-10}$ (blue), $10^{-12}$ (orange) and $10^{-15}$ (green). Results are shown for assuming a strong harmonic trap along the $z$ direction and for $l_z$ choices of a) $l_z=0.1$, b)$l_z=1$ and c) $l_z=5$.}
    \label{fig:dipole_harmonic_dims}
\end{figure}

We again construct the above interaction kernel in momentum space and decompose into MPS form, for various truncation cutoffs and $lz$ shown in Fig.~\ref{fig:dipole_harmonic_dims}. We again observe the same logarithmic scaling with MPS length as was observed in the main text.

\section{Dipolar interaction truncation}
\label{App:momentum_cutoff}
Constructing the interaction kernel as in Eq.~\eqref{eq:app_phi} with the Fourier transform results in a periodic interaction term, and hence the interaction term technically includes the effect of interactions with identical copies of the BEC outside of the original simulation domain. In most cases, this will not provide a substantial impact on the physics, however one may desire to truncate the long range interaction as follows, 
\begin{align}
 U_\mathrm{dd}^\mathbf{R}(\mathbf{r})=\left\{\begin{array}{ll}
                  U_\mathrm{dd}(\mathbf{r}), & |\mathbf{r}|<|\mathbf{R}|,\\[0.1cm]
                  0, & \mathrm{else}
                 \end{array}\right.
\end{align}
where $\mathbf{R}$ can be viewed as the truncation radius, beyond which we ignore all interactions. We therefore expect that as $\mathbf{R}$ tends to infinity, we recover the fully periodic treatment. The form of this truncation is presented in Ref.~\cite{Smith_2023} for three dimensions. However, here we restrict to two dimensions by again assuming a uniform condensate along the $z$ axis. The radius truncated kernel can then be given by
\begin{equation}
    \tilde{U}_\mathrm{dd}^R(\mathbf{k}) = \frac{1}{2} +3\frac{k_y^2 -k_x^2}{k^2} [\frac{1}{2} - \frac{J_1(\mathbf{k}R)}{\mathbf{k}R}] \ ,
\end{equation}
with $J_1(\cdot)$ being the Bessel function of the first kind. 

\begin{figure}[htb!]
    \centering
    \includegraphics[width=\linewidth]{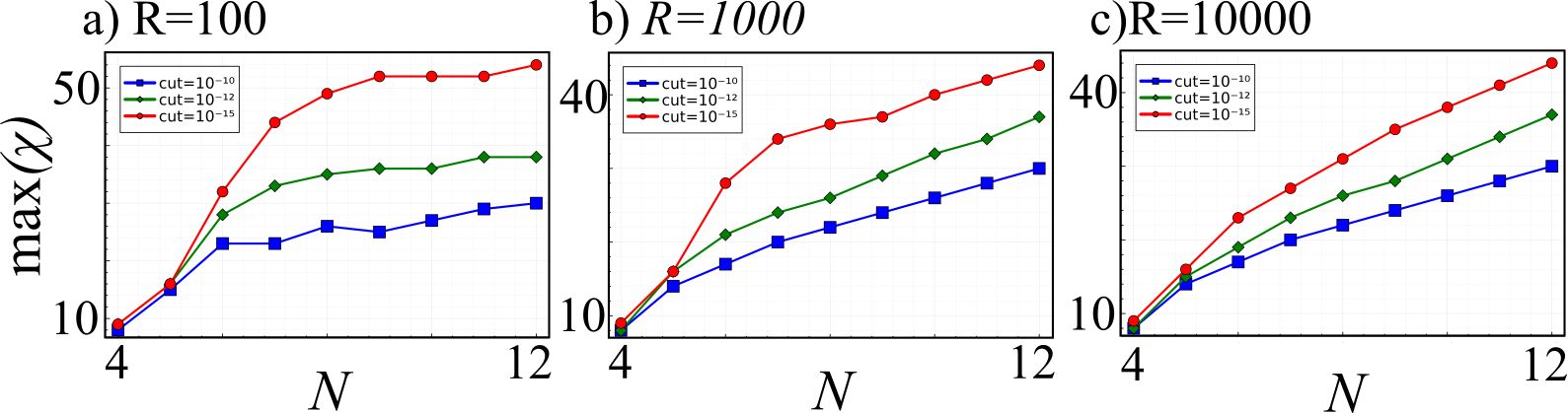}
    \caption{Bond dimension scaling for various truncation radii $R$. As we increase $R$, the bond dimension scale more akin to the periodic case of $R=\infty$.}
    \label{fig:app_trunc_Udd}
\end{figure}

We construct this interaction potential as an array in momentum space, before performing successive SVDs to put into MPS form. The resultant bond dimension scalings for various truncations and interaction Radii $\mathbf{R}$ are illustrated in Fig.~\ref{fig:app_trunc_Udd}. We observe that the resultant bond dimensions are larger for smaller truncation radii, and still grow with system size. It can be seen that as $R\rightarrow \infty$, we recover the same scaling as the fully periodic treatment in Fig.~\ref{fig:dipole_dims}.

\end{document}